%% file: ms_resubmit_arxiv_v2.tex
\documentclass{emulateapj}
\usepackage{epsfig,graphics}
\usepackage[colorlinks=true,
  citecolor=red,  
  pagecolor=blue, 
  linkcolor=blue, 
  menucolor=blue, 
  urlcolor=blue,  
  linkbordercolor={0 0 1},
  frenchlinks=True,
  breaklinks]{hyperref}

\begin{document}
\submitted{Received 2010 August 31; accepted 2010 November 10; published 2010 December 21}

\newcommand{\Ha}{H$\alpha$}
\newcommand{\OIII}{[\ion{O}{3}]}
\newcommand{\NII}{[\ion{N}{2}]}
\newcommand{\OII}{[\ion{O}{2}]}
\newcommand{\SIII}{[\ion{S}{3}]}
\newcommand{\SII}{[\ion{S}{2}]}
\newcommand{\C}{$\kappa({\rm L})$}
\newcommand{\mm}{$\mu$m}
\newcommand\aper[1]{#1\arcsec\ diameter}%

\newcommand{\Rcf}{R_{\rm C}}
\newcommand{\Rc}{$\Rcf$}

\newcommand{\zphotf}{z_{\rm phot}}
\newcommand{\zspecf}{z_{\rm spec}}
\newcommand{\zphot}{$\zphotf$}
\newcommand{\zspec}{$\zspecf$}

\newcommand{\Hb}{H$\beta$}
\newcommand{\Msun}{$M_{\sun}$}

\newcommand{\phistar}{\Phi_{\star}}
\newcommand{\Lstar}{L_{\star}}
\newcommand{\SFRd}{\rho_{\rm SFR}}
\newcommand{\iyr}{yr$^{-1}$}
\newcommand{\vMpc}{Mpc$^{-3}$}

\newcommand{\fluxunit}{erg s$^{-1}$ cm$^{-2}$}

\newcommand{\mumod}{\langle\log{({\rm EW_0}/{\rm \AA})}\rangle}
\newcommand{\sigmamod}{\sigma[\log{({\rm EW}/{\rm \AA})}]}

\title{The \Ha\ Luminosity Function and Star Formation Rate Volume Density at
  $z=0.8$ from the NEWFIRM \Ha\ Survey}

\author{Chun Ly,\altaffilmark{1,2,6,7} Janice C. Lee,\altaffilmark{2,8} Daniel A. Dale,\altaffilmark{3}
  Ivelina Momcheva,\altaffilmark{3} Samir Salim,\altaffilmark{4} Shawn Staudaher,\altaffilmark{3} 
  Carolynn A. Moore,\altaffilmark{3} and Rose Finn\altaffilmark{5}}
\shorttitle{\Ha\ Luminosity Function and SFR Volume Density at $z=0.8$}
\shortauthors{Ly et al.}
\altaffiltext{1}{Department of Physics and Astronomy, UCLA, Los Angeles, CA, USA; chunly@stsci.edu}
\altaffiltext{2}{Carnegie Observatories, Pasadena, CA, USA}
\altaffiltext{3}{Department of Physics and Astronomy, University of Wyoming, Laramie, WY, USA}
\altaffiltext{4}{Department of Astronomy, Indiana University, Bloomington, IN, USA}
\altaffiltext{5}{Department of Physics, Siena College, Loudonville, NY, USA}
\altaffiltext{6}{Current address: Space Telescope Science Institute, Baltimore, MD, USA.}
\altaffiltext{7}{Giacconi Fellow.}
\altaffiltext{8}{Carnegie Starr Fellow.}

\newcommand{\zlo}{0.80}
\newcommand{\zhi}{2.19}
\newcommand{\NBlo}{1.18}
\newcommand{\NBhi}{2.09}

\newcommand{\alphafa}{-1.6}           
\newcommand{\lstarfa}{43.03\pm0.17}   
\newcommand{\phistarfa}{-3.20\pm0.13} 
\newcommand{\LDfa}{40.17\pm0.05}      

\newcommand{\LDflima}{40.04\pm0.05}   
\newcommand{\SFRfa}{-0.93\pm0.05}     
\newcommand{\SFRfAGNa}{-0.98\pm0.05}  
\newcommand{\SFRflima}{-1.06\pm0.05}  
\newcommand{\alphaf}{$\alpha = \alphafa$} 
\newcommand{\lstarf}{$\Lstar = 10^{\lstarfa}$ erg s$^{-1}$} 
\newcommand{\phistarf}{$\phistar = 10^{\phistarfa}$ \vMpc} 
\newcommand{\LDf}{$\mathcal{L} = 10^{\LDfa}$ erg s$^{-1}$ \vMpc} 
\newcommand{\LDflim}{$\mathcal{L} = 10^{\LDflima}$ erg s$^{-1}$ \vMpc} 
\newcommand{\SFRf}{$\SFRd = 10^{\SFRfa\pm0.04}$ \Msun\ \iyr\ \vMpc} 
\newcommand{\SFRfAGN}{$\SFRd = 10^{\SFRfAGNa\pm0.04}$ \Msun\ \iyr\ \vMpc} 
\newcommand{\SFRflim}{$\SFRd = 10^{\SFRflima\pm 0.05}$ \Msun\ \iyr\ \vMpc} 

\newcommand{\LDbina}{40.08\pm0.03}   
\newcommand{\LDbinb}{40.01\pm0.04}   
\newcommand{\LDfbina}{$\mathcal{L} = 10^{\LDbina}$ erg s$^{-1}$ \vMpc} 
\newcommand{\LDfbinb}{$\mathcal{L} = 10^{\LDbinb}$ erg s$^{-1}$ \vMpc} 

\newcommand{\alphapa}{-1.6\pm0.19}    
\newcommand{\lstarpa}{43.00\pm0.52}   
\newcommand{\phistarpa}{-3.20\pm0.54} 
\newcommand{\LDpa}{40.15\pm0.18}      
\newcommand{\LDplima}{40.01\pm0.08}   
\newcommand{\SFRpa}{-0.96\pm0.18}     
\newcommand{\SFRpAGNa}{-1.00\pm0.18}  
\newcommand{\SFRplima}{-1.10\pm0.08}  
\newcommand{\alphap}{$\alpha=\alphapa$} 
\newcommand{\lstarp}{$\Lstar=10^{\lstarpa}$ erg s$^{-1}$}
\newcommand{\phistarp}{$\phistar = 10^{\phistarpa}$ \vMpc}
\newcommand{\LDp}{$\mathcal{L} = 10^{40.05\pm0.13}$ erg s$^{-1}$ \vMpc}
\newcommand{\LDplim}{$\mathcal{L} = 10^{\LDplima}$ erg s$^{-1}$ \vMpc} 
\newcommand{\SFRp}{$\SFRd = 10^{\SFRpa\pm0.04}$ \Msun\ \iyr\ \vMpc} 
\newcommand{\SFRpAGN}{$\SFRd = 10^{\SFRpAGNa\pm0.04}$ \Msun\ \iyr\ \vMpc} 
\newcommand{\SFRplim}{$\SFRd = 10^{\SFRplima\pm 0.05}$ \Msun\ \iyr\ \vMpc} 

\newcommand{\LDbin}{$\mathcal{L} = 10^{40.08\pm0.07}$ erg s$^{-1}$ \vMpc} 
\newcommand{\Llim}{$1.0\times10^{41}$ erg s$^{-1}$}    
\newcommand{\Llimext}{$3.0\times10^{41}$ erg s$^{-1}$} 
\newcommand{\AGNf}{11\%}

\newcommand{\VLs}{42.97\pm0.27}
\newcommand{\VPs}{-2.76\pm0.32}
\newcommand{\Va}{-1.34\pm0.18}

\newcommand{\SLs}{42.33^{+0.16}_{-0.12}}
\newcommand{\SPs}{-2.51^{+0.17}_{-0.16}}
\newcommand{\Sa}{-1.64\pm0.21}

\newcommand{\NHa}{522}  
\newcommand{\NNB}{1218} 
\newcommand{\NHas}{394} 
\newcommand{\NNBs}{818} 

\newcommand{\Pspec}{46\%}  
\newcommand{\Pspecs}{62\%} 

\begin{abstract}
We present new measurements of the H$\alpha$ luminosity function (LF) and star formation rate (SFR) volume density for
galaxies at $z\sim0.8$. Our analysis is based on \NBlo\mm\ narrowband data from the NEWFIRM H$\alpha$ (NewH$\alpha$)
Survey, a comprehensive program designed to capture deep samples of intermediate redshift emission-line galaxies using
narrowband imaging in the near-infrared. The combination of depth ($\approx1.9\times10^{-17}$ \fluxunit\ in H$\alpha$ at
3$\sigma$) and areal coverage (0.82 deg$^2$) of the \NBlo\mm\ observations complements other recent H$\alpha$ studies
at similar redshifts, and enables us to minimize the impact of cosmic variance and place robust constraints on the
shape of the LF. The present sample contains \NNBs\ NB118 excess objects, \NHas\ of which are selected as H$\alpha$
emitters. Optical spectroscopy has been obtained for \Pspecs\ of the NB118 excess objects. Empirical optical broadband
color classification is used to sort the remainder of the sample. A comparison of the LFs constructed for the four
individual fields covered by the observations reveals significant cosmic variance, emphasizing that multiple, widely
separated observations are required for such analyses. The dust-corrected LF is well described by a Schechter
function with \lstarp, \phistarp, and \alphap. We compare our H$\alpha$ LF and SFR density to those at $z\lesssim1$,
and find a rise in the SFR density $\propto(1+z)^{3.4}$, which we attribute to significant $\Lstar$ evolution. Our
H$\alpha$ SFR density of $10^{-1.00\pm0.18}$ \Msun\ \iyr\ \vMpc\ is consistent with UV and \OII\ measurements at
$z\sim1$. We discuss how these results compare to other H$\alpha$ surveys at $z\sim0.8$, and find that the different
methods used to determine survey completeness can lead to inconsistent results. This suggests that future surveys
probing fainter luminosities are needed, and more rigorous methods of estimating the completeness should be adopted
as standard procedure (for example, with simulations which try to simultaneously reproduce the observed H$\alpha$ LF
and equivalent width distributions).
\end{abstract}



\section{INTRODUCTION}\label{0}
\indent The luminosity of the \Ha\ nebular emission line is a star formation rate (SFR) indicator valued for its
relatively direct physical connection to short-lived massive stars.  In the local universe, it has been well calibrated
\citep[e.g.,][]{kennicutt98,kennicutt09}, and extensively used to measure the SFRs of individual galaxies, as well as
the SFR density over cosmic volumes
\citep[e.g.,][]{kennicutt83,gallego95,salzer01,gavazzi02,brinchmann04,hanish06,meurer06,ly07,dale08,kennicutt08,lee09}. 

While it is desirable to extend \Ha\ studies of galaxies to earlier cosmic times, such work is observationally difficult
because \Ha\ is redshifted into the infrared beyond $z\sim0.4$. Early attempts yielded samples of \Ha\ emitters that
were small in size and did not sample representative cosmic volumes, because of the limited depth and areal coverage of
the observations \citep[e.g.,][]{yan99,hopkins00,tresse02}. Other SFR indicators, which are accessible in the optical
at higher redshift (e.g., the rest-frame UV continuum or other bluer emission lines), have therefore been more commonly
used. Measurements of the SFR density have now been made at redshifts as high as $\sim6$. A sharp, order of magnitude
decline is seen in the star formation activity of the universe from $z\sim1$ to the present day
\citep[see e.g.,][]{hopkins04}. However, the amalgam of measurements that constitute our current understanding of the
cosmic star formation history shows a scatter of at least a factor of a few, which considerably reduces their usefulness
as a constraint on models of galaxy evolution.

In this context, it is important to determine the extent to which systematics between different SFR indicators (such as
the variable impact of dust attenuation and dependence on metallicity) contribute to the observed scatter in the cosmic
star formation history. In particular, it is useful to trace the history with a consistent indicator throughout time,
and then compare the overall histories determined with different indicators.  This paper represents a step in this
process, and aids in the robust extension of \Ha\ measurements of the SFR density to higher redshift. 

Here, we present \Ha\ luminosity functions (LFs) and SFR densities at $z=\zlo$.  At this redshift, the age of the universe is
6.6 Gyr, assuming a [$\Omega_{\Lambda}$, $\Omega_M$, $h_{70}$] = [0.7, 0.3, 1.0] cosmology, which we adopt throughout.
Our analysis is based on narrowband (NB) observations from the NEWFIRM \Ha\ (New\Ha) Survey
(J. C. Lee et al. 2011, in preparation), which has been conducted
with the NOAO Extremely Wide-Field Infrared Imager \citep[NEWFIRM;][]{probst04,probst08} at the KPNO 4 m telescope.
The current sample contains $\sim$400 \Ha\ emitting galaxies above 3$\sigma$ significance, which have been identified
over an area of 0.82 deg$^2$.

This paper is organized as follows. In Section \ref{1}, we give an overview of the New\Ha\ data that are used in our analysis.
In Section \ref{sec3}, we describe the selection of NB118 excess emitters. We also discuss the techniques (i.e., dedicated
follow-up spectroscopy and empirical broadband color classification) that are used to identify the emission line(s)
responsible for the narrowband excess.
A description of how we compute emission-line fluxes, luminosities, and equivalent widths (EWs) is provided
in Section \ref{2.3}. In Section \ref{3}, estimates of the survey's completeness and the surveyed volume are presented, and
Section \ref{4} presents the \Ha\ LF and the comoving SFR density at $z\sim0.8$. Comparisons of these
results with existing \Ha\ measurements are described in Section \ref{5}. We
also compare our results with that of other recent $z\sim0.8$ narrowband \Ha\ surveys, and examine the reasons why
inconsistent results may arise between these studies.
A discussion of our results and their implications for the evolution of typical galaxies at $z\sim0.8$ are provided in
Section \ref{6}, and concluding remarks are provided in Section \ref{8}. Magnitudes are reported on the AB system \citep{oke74}.


\input tab1_arxiv.tex

\section{Observations and Data Reduction}\label{1}


\subsection{The New\Ha\ Survey}
The New\Ha\ Survey is an ongoing campaign designed to extend deep, wide searches for emission-line
galaxies into the intermediate redshift universe. The survey takes advantage of the 27\farcm6 $\times$ 27\farcm6
field of view of the NEWFIRM, which achieved first light on the KPNO 4m
telescope in 2007 February, and became available for general observing in 2007 November.

New\Ha\ uses narrowband observations to identify emission-line galaxy candidates. Imaging is taken through 1\% filters
that are designed to sample low OH airglow windows in the near-infrared, and objects are selected by detection of a
photometric excess in a narrowband, relative to continuum measurements in a broadband. Bandpasses are centered at
\NBlo\ $\mu$m and \NBhi\ $\mu$m, which capture \Ha\ emission at redshifts of \zlo\ and \zhi, respectively. The
continuum flux is constrained with $J$ and $K_s$ imaging. A combination of techniques, including dedicated follow-up
spectroscopy and empirical broadband color classification, is used to identify the emission line(s) responsible for
the narrowband excess, as discussed in more detail below. A full description of the overall survey is provided in
J. C. Lee et al. (2011, in preparation). Here, we give a summary of the \NBlo\ $\mu$m imaging observations, data
reduction, and selection technique used to construct a robust sample of \Ha\ emitters at $z=\zlo$. 


\subsection{NEWFIRM Observations}\label{1.1}
The analysis presented in this paper is based upon NEWFIRM
$J$ band\footnote{$\lambda=1.250$ \mm; $\delta\lambda=0.180$ \mm} and \NBlo\
$\mu$m narrowband (hereafter NB118)\footnote{$\lambda=1.184$ \mm; $\delta\lambda=0.011$ \mm} observations of
areas in the Subaru-XMM Deep Survey \citep[SXDS;][]{furusawa08} and Cosmic Evolution Survey \citep[COSMOS;][]{scoville07}
extragalactic deep fields. The observations were carried out
in 2007 December, 2008 September, and 2008 October at the KPNO 4m telescope.  Data were acquired for three regions
in the SXDS and one region in COSMOS, each spanning the $\sim750$ arcmin$^2$ subtended by the detector array of the
camera.  In total, these observations cover 0.82 deg$^{2}$ for a comoving volume of
$9.12\times10^4$ $h_{70}^{-3}$ Mpc$^3$ at $z=0.8$.

We follow standard near-infrared observing procedures to obtain the data. Integration times of 240 and 30 s are used
for the individual NB118 and $J$ exposures, respectively. The NB118 exposures are read-out with eight Fowler samples.
In 2008, on board co-adding was enabled in the camera, and was used for our $J$-band observations, with every two
$J$ exposures being summed. A combination of 9-, 6-, and 4-point dither patterns is used to cover a 75\arcsec\ square
grid with 61 positions. The dithers serve to smooth over cosmetic defects, bridge the 35\arcsec\ gaps between
NEWFIRM's four 2048 $\times$ 2048 InSb arrays, and enable the rejection of pixels affected by persistent afterimages
caused by the latency properties of the detectors. The pattern, with small offsets from the initial position, is
repeated as necessary to achieve a minimum 3$\sigma$ depth of 23.5 AB (NB118) and 23.7 AB ($J$), in apertures
containing at least $\sim$80\% of the flux of a point source, given the seeing conditions during the observations
(see Section \ref{sec:NB118select}). Cumulative integration times range between 8.2 and 12.7 hr in NB118 and 2.3
and 4.0 hr in $J$.  The median seeing during our observations was $\sim1\farcs2$, and varied between 1\farcs0 and 1\farcs9,
hence point sources are adequately sampled with 0\farcs4 pixels. Sky conditions were mostly photometric, but some data
were taken through thin cirrus. A summary of the observations is given in Table~\ref{table1}.


\subsection{NEWFIRM Data Reduction}\label{1.2}

Data reduction is performed with our own automated PyRAF-based pipeline, which builds upon routines from the
IRAF/{\sc nfextern} package and is optimized for the processing of New\Ha\ observations. Our procedures follow
standard, iterative, near-infrared reduction techniques to produce flat-fields, subtract the sky background, and
reject artifacts, particularly those due to persistent afterimages from bright sources.

The New\Ha\ pipeline processes the data in two passes.  One of the purposes of the initial pass is to create a
deep object mask that is used to construct flat-fields and sky frames from the science images themselves. 
The pipeline first subtracts the dark current, masks for bad pixels, corrects for the nonlinear response of the
detector, and performs a preliminary sky-subtraction using a median of the temporally closest five exposures.
The geometric distortion is rectified, and the astrometry is calibrated relative to the Two Micron All Sky
Survey (2MASS) catalog. The
dithered science images are then projected onto a common pixel grid and stacked. A deep object mask is made from
this initial stack, and applied to the individual science exposures on the original pixel grid.

The first pass sky-subtracted frames are also used to identify artifacts due to persistent afterimages. Object masks
are created for each {\it individual} frame, and pixels in a given image are compared to corresponding pixels from
previous frames. Pixels containing object flux in consecutive frames are masked. Non-science images, such as short
exposures used to check and adjust the telescope pointing, are also included in this process.  This method  
flags roughly 95\% of visually apparent persistent artifacts. The remaining 5\% are faint and decay quickly enough
in subsequent frames that they are not detectable in the final mosaics after the stacking of the entire data set. 

Flat-fields for each night are made by combining $J$-band science images, which have been masked for both real objects
and persistent sources. The flats are used to normalize the response within each detector in both the $J$ and NB118
images. Corrections that account for sensitivity variation between detectors are also applied.

In the second pass, all the data are flat-fielded, and the sky subtraction is again performed, but with stacks of
temporally neighboring frames which have now been masked of all sources.  Any problematic frames (e.g., due to 
read-out problems or jumps in the telescope pointing) are rejected.  The flat-fielded, sky-subtracted,
persistence-masked frames are then combined to produce the final mosaic. The astrometry of the mosaics is tested
against 2MASS sources, as well as against sources in the COSMOS optical catalog, and is found to be accurate to within 
0\farcs15--0\farcs20 with negligible systematic offsets. 

Photometric calibration of the mosaics is performed using 150--300 unsaturated 2MASS \citep{skrutskie06} sources in
each field. We check for systematic errors as a function of both $J$/NB118 magnitude and radius from the mosaic center, 
and find no significant offsets. The resultant zeropoints are accurate to within 0.05 mag. Absolute flux calibration
for the 2MASS catalog is based on Vega \citep{kurucz79}, hence we convert to AB magnitudes by convolving the
filter bandpasses with the Vega spectrum, and find that $m({\rm AB})-m({\rm Vega}) = 0.87$ (NB118) and 0.95 ($J$).


\section{Sample Selection}\label{sec3}

\begin{figure*} 
  \epsscale{0.45}
  \plotone{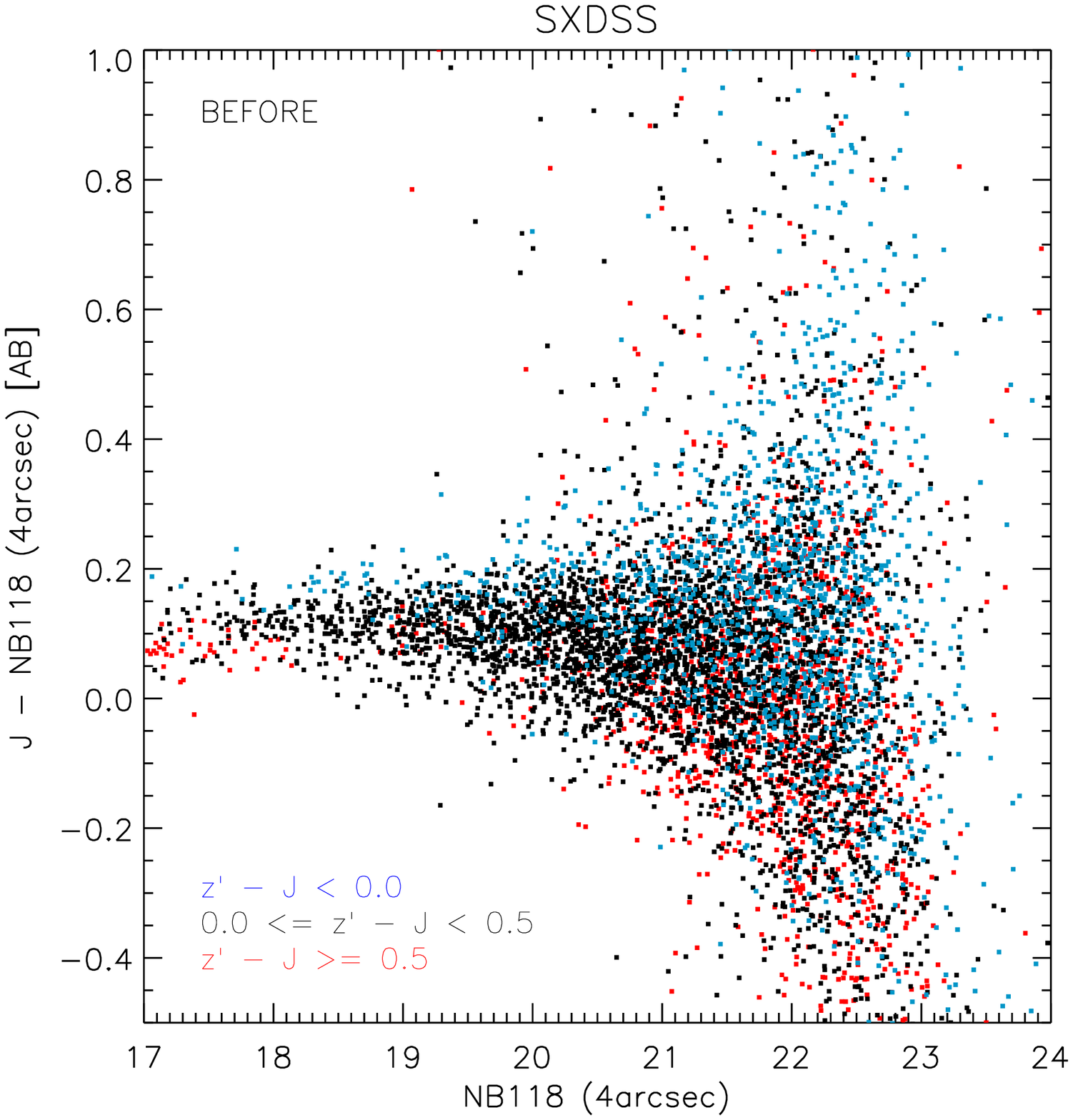} \plotone{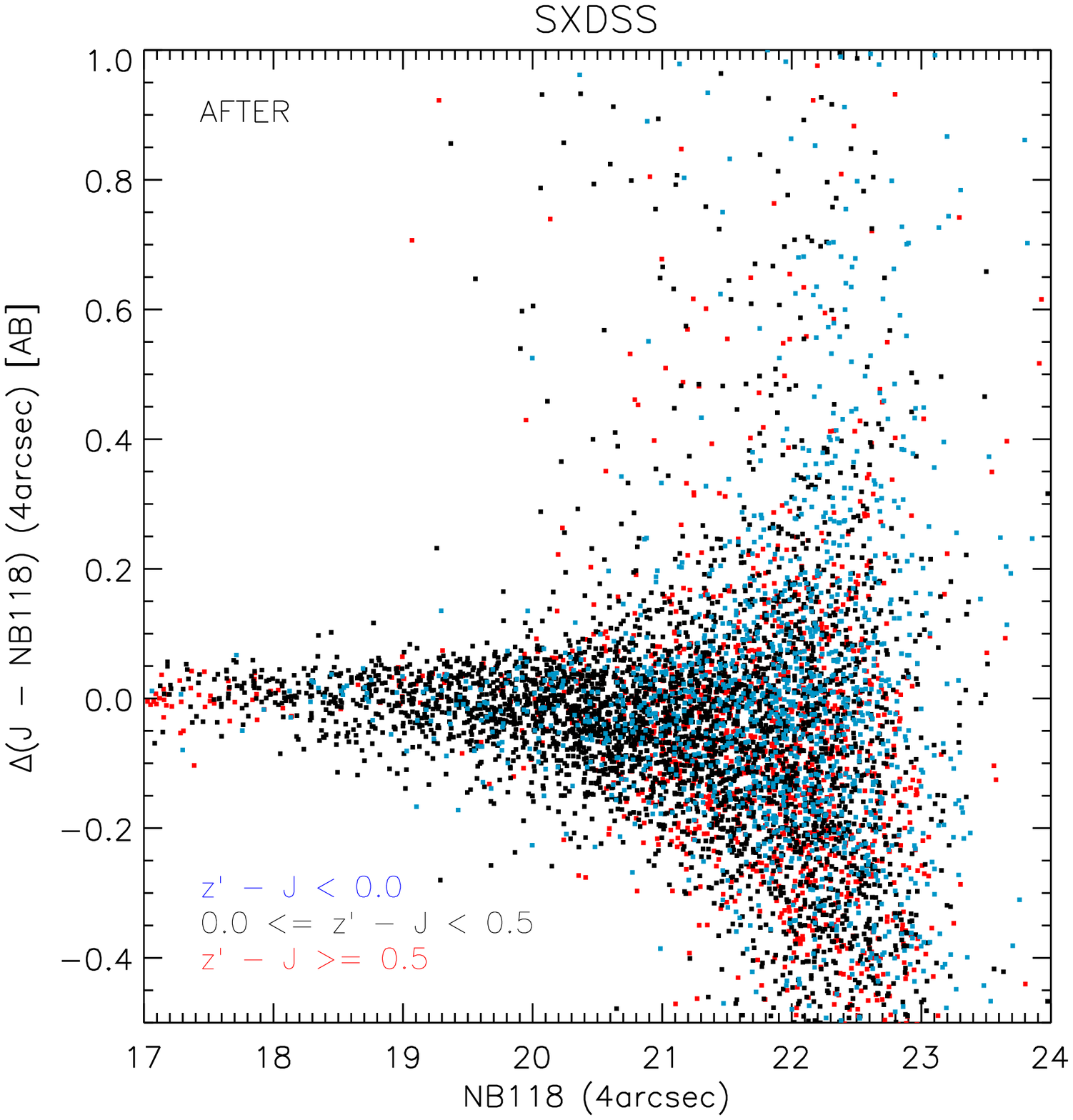}
  \caption{Left: standard broadband--narrowband color magnitude selection diagrams, which illustrate the
    dependence of the mean value of $J-{\rm NB118}$ for pure continuum sources ``$\langle J-{\rm NB118} \rangle_c$''
    on the slope of the continuum, as measured by $z^\prime - J$. The points are color coded to correspond to
    different ranges of $z^\prime - J$, as indicated in the plot. Sources with redder continua have lower
    $\langle J-{\rm NB118} \rangle_c$ compared with those with bluer continua. Right: the color magnitude
    diagram after the dependence on $z^\prime - J$ has been removed, and $\langle J-{\rm NB118} \rangle_c$ is
    subtracted. The resultant $\Delta(J-{\rm NB118})$ is the color (or narrowband) excess used to compute
    emission-line fluxes and equivalent widths. (A color version of this figure is available in the online journal.)}
  \label{fig:zJ}
\end{figure*}


\begin{figure*} 
  \epsscale{0.9}
  \plotone{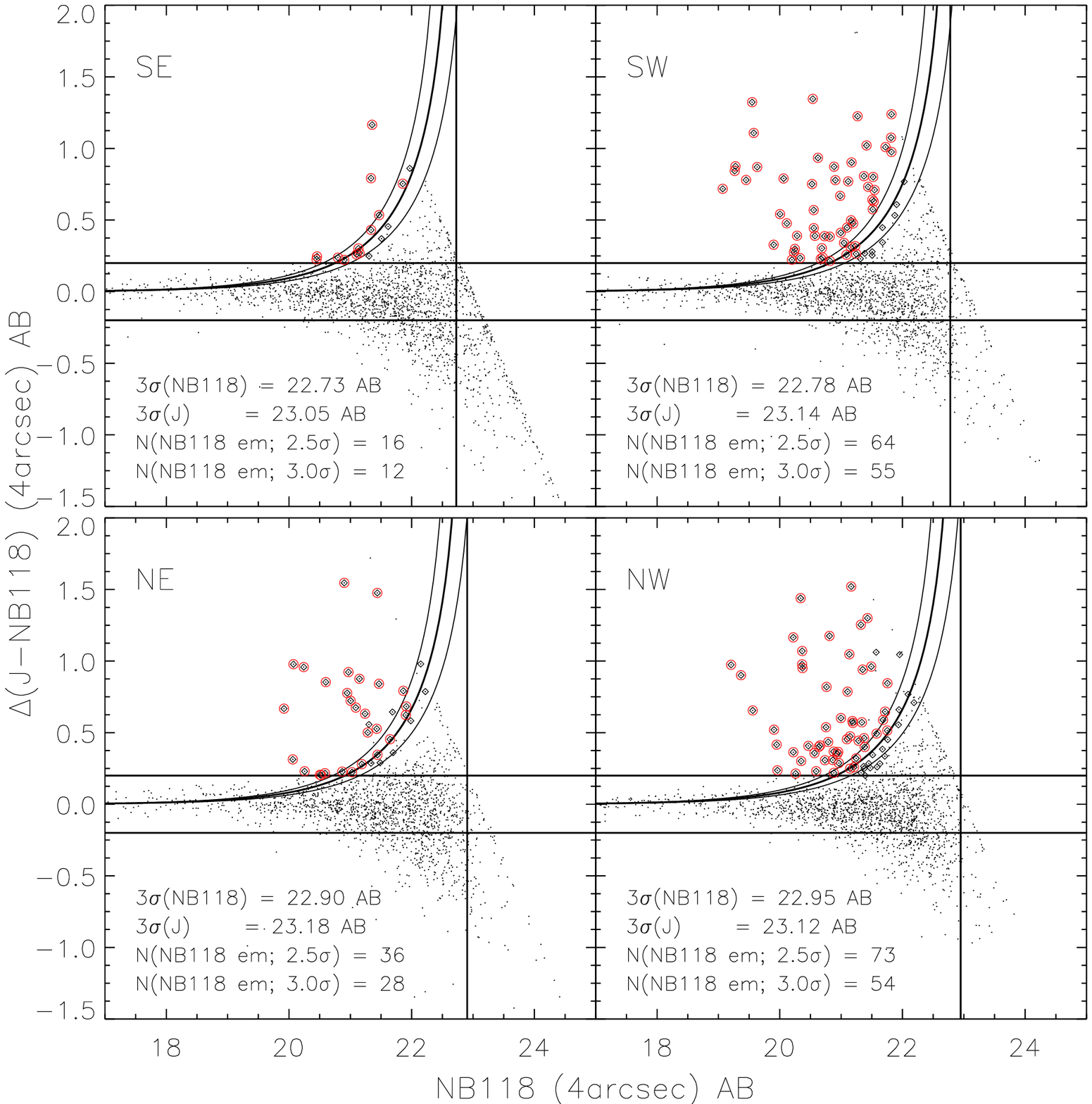}
  \caption{Color--magnitude diagrams illustrating the selection of NB118 excess emitters in the SXDS-S, where the
    photometry has been measured in the larger choice of aperture (4\arcsec).  Note that the color excess
    $\Delta(J-{\rm NB118})$ (see Equations~(\ref{eq:colorexcess1})--(\ref{eq:colorexcess2})) rather than raw
    $J-{\rm NB118}$ color is plotted.  The horizontal lines indicate $\pm$0.2 mag from zero excess; the minimum 
    accepted $\Delta(J-{\rm NB118})$ is 0.2 mag.  The vertical lines indicate the 3$\sigma$ depth of the NB118
    photometry. The three curves shown correspond to the {\it average} values of the color excess at 3.0$\sigma$, 2.5$\sigma$,
    and 2.0$\sigma$ significance, as determined from measurement of the image background with random apertures,
    and Gaussian fitting of the resultant distribution. Sources with $\Delta(J-{\rm NB118})$ values significant
    at the 3$\sigma$ level are highlighted with red circles, while those significant at the 2.5$\sigma$ level
    are enclosed in black diamonds. The selection is done on a per-quadrant basis, to account for the varying
    sensitivities of the detectors.}  
  \label{fig:colormag}
\end{figure*}



\subsection{Selection of Narrowband Excess Emitters}\label{sec:NB118select}
Sources that show a significant $J-$ NB118 color excess are selected as emission-line object candidates. Our
selection procedure follows general techniques commonly used in narrowband surveys
\citep[e.g.,][]{fujita03,ly07,shioya08,villar08,sobral09}. 

To construct our sample, we first use SExtractor \citep[version 2.5.0;][]{bertin96} in dual-image mode to generate
source catalogs for each field. That is, sources are identified on the NB118 image, and fluxes are measured on both
$J$ and NB118 images in matched apertures at the position of every 3$\sigma$ NB118 detection. Detection in the
$J$ band is not required for inclusion in the candidate list.
 
In each field, photometry is performed in two sets of apertures. The sizes of the apertures for each pair of $J$ and
NB118 images are chosen to contain at least 80\% and 99\% of light from a point source, given the size of the
point-spread function
(Table~\ref{table1}) and assuming a Gaussian profile. In the SXDS-N and SXDS-W fields, the seeing during our
observations was between 1\farcs1 and 1\farcs2, and thus 2\arcsec\ and 3\arcsec\ diameter apertures are used. The
aperture sizes are increased to 2\farcs5 and 4\arcsec\ for the SXDS-S and COSMOS fields, where the seeing was worse
and varied from 1\farcs25 to 1\farcs6. One of the purposes of using two aperture sizes is to allow the selection to
be sensitive to galaxies with concentrated star formation as well as those with more extended nebular emission. In
order to derive global quantities, however, the \Ha\ luminosities and EWs for all selected sources are
calculated from fluxes measured in the larger aperture.

Sources are considered to have a significant narrowband excess if:\\
\noindent
1. They have a color above a minimum threshold given by
\begin{equation}
\Delta(J-{\rm NB118}) \geq 0.2\; \mbox{mag,}
\label{eq:colorexcess1}
\end{equation}
where $\Delta(J-{\rm NB118})$ is defined below, and 0.2 mag roughly corresponds to the $5\sigma$ scatter in
$\Delta(J-{\rm NB118})$ for bright point sources with 18 $<{\rm NB118}<$ 20 mag. It is imposed to help exclude bright 
foreground sources with blue continua from the candidate list.

\noindent
2. The color is significant at the 3$\sigma$ level 
\begin{equation}
\Delta(J-{\rm NB118}) \geq 3 \sqrt{\sigma_{J}^2+\sigma_{\rm NB118}^2}\;.
\label{eq:colorexcess2}
\end{equation}
Other recent near-infrared narrowband surveys have adopted a lower threshold of 2.5$\sigma$
\citep[hereafter V08 and S09]{villar08,sobral09}.\defcitealias{villar08}{V08}\defcitealias{sobral09}{S09} 
We also extract a secondary sample with this relaxed significance criterion to facilitate comparisons with these studies
(Section \ref{5}).

Two issues critical to the robust application of these criteria for the selection of line emitters are the determination
of $\Delta(J-{\rm NB118})$, and the calculation of accurate photometric errors.  We discuss each of these in turn.

Flux excess in the NB118 filter cannot be directly calculated from the $J-{\rm NB118}$ color since the NB118 filter
does not fall at the center of the $J$ bandpass, but rather was designed to sample the low-OH airglow window toward
the blue edge.  This not only results in a non-zero mean color for pure continuum sources
$\langle J-{\rm NB118} \rangle_c$, but also leads to variation about the mean color as a function of continuum slope.  
This systematic is illustrated in the left panel of Figure~\ref{fig:zJ}, which shows a standard color--magnitude
selection diagram for one of our fields. The points in the diagram are color-coded to correspond to different ranges
in the $z^\prime-J$ color, which provides a measure of the continuum slope.  Sources with red continua have lower values
of $\langle J-{\rm NB118} \rangle_c$ compared with those with blue continua, as expected.  Here, the $J$ photometry is
from our NEWFIRM observations, while the $z^\prime$ measurements are from Subaru/Suprime-Cam (see \citealt{furusawa08}
for SXDS and \citealt{taniguchi07} for COSMOS).

To account for this systematic, we define 
\begin{equation}
\Delta(J-{\rm NB118})=(J-{\rm NB118})-\langle J-{\rm NB118} \rangle_c
\end{equation}
where
\begin{equation}
\langle J-{\rm NB118} \rangle_c=f(z^\prime-J).
\end{equation}
A linear fit is performed for sources with $J-{\rm NB118}$ within $\pm$0.25 mag of the overall mean value. We find that
$\langle J-{\rm NB118} \rangle_c = -0.15(z^\prime-J) + {\rm const.}$ for $z^\prime-J<0.5$. For $z^\prime-J>0.5$, there is
no apparent slope so a constant value is assumed for redder sources.  Accordingly, we compute corrections and apply them
relative to the value of $\langle J-{\rm NB118} \rangle_c$ for sources with an intermediate $z^\prime-J$ of 0.25 mag.
The right panel of Figure~\ref{fig:zJ} shows the color--magnitude diagram after this correction is applied and
demonstrates the removal of the systematic.

Photometric uncertainties are calculated by combining the errors generated by SExtractor with measurements of the
background noise taken through a large number of apertures randomly placed on the images. This hybrid scheme ensures
that our uncertainties account for (1) correlated (non-Poissonian) noise arising from pixel interpolation during
astrometric reprojection in the image reduction (captured in the random aperture measurements) and (2) local
variations in the noise, for example, due to non-uniformity in the exposure map (captured by the SExtractor errors).
A detailed discussion of these issues is given in \cite{gawiser06}.

The global, average uncertainty due to the background can be robustly measured using a large number of apertures
(identical in size to the apertures used for the photometry) that randomly sample the sky in the images.  We perform
this exercise separately for different quadrants in the image to account for quantum efficiency differences between the 
NEWFIRM detectors.\footnote{In Table~\ref{table1}, note that while three detectors have comparable sensitivities, 
the fourth is less sensitive by about 0.3 mag. The primary cause is an extra layer of anti-reflection coating, which
was inadvertently applied to the detector.} The distribution of measurements is well fitted by a Gaussian function, and
the standard deviation given by the fit provides the 1$\sigma$ uncertainty. The average depths of our observations
determined in this way are given in Table~\ref{table1}.

To incorporate information on local variations in the noise that is included in the SExtractor errors, the SExtractor
estimates are compared with those computed from the random apertures.  We find that SExtractor yields lower
uncertainties, which is expected since the program assumes that the noise is purely Poissonian and simply computes the
error from the global pixel-to-pixel rms. To account for the contribution of non-Poissonian noise components, we scale
the median SExtractor error (as a function of the object flux and on a per-quadrant basis) to match the average
uncertainty determined from the random apertures.
The errors from SExtractor are increased by 2\%--51\%, depending upon the aperture size used.
These scaled SExtractor errors are adopted to evaluate the significance of the color excess detected in our observations.

The selection criteria are illustrated in Figure~\ref{fig:colormag}, which shows the color--magnitude diagrams for one
of our fields, where the photometry has been measured in the larger choice of aperture. Horizontal lines indicating the
minimum accepted color excess are plotted, along with curves representing the average values of the color excess at
3.0$\sigma$, 2.5$\sigma$, and 2.0$\sigma$ significance (as determined using random apertures). Objects selected at 3$\sigma$ (red circles)
and 2.5$\sigma$ (black diamonds) are both shown.  Hereafter, quantities related to the 2.5$\sigma$ samples are quoted
in brackets unless otherwise noted.

The selection is performed separately for the photometry determined through the two sets of apertures, as well as for
the individual quadrants in each field. The use of a second, larger detection aperture results in a $\sim$10\% increase
of the number of narrowband excess candidates. In total, we find 150--300 [250--450] candidates per field. All candidates
are inspected by visual examination of the NEWFIRM $J$ and NB118 data, alongside publically available deep $z\arcmin$
imaging. The inspection process leads to the removal of four sources from the sample (two are artifacts and two are
significantly blended with adjacent candidates). The overall process yields a total sample of \NNBs\ [\NNB] NB118
excess emitters in the combined survey area of 0.82 deg$^{2}$. A summary of the number of sources identified in each
field is given in Table~\ref{table2}. A catalog of NB118 excess emitters, including emission-line fluxes and
EWs, will be provided in a forthcoming paper (J. C. Lee et al. 2011, in preparation).

\input tab2.v1.arxiv.tex


\subsection{Identification of \Ha\ Emitters}\label{2.1}
An excess in the narrowband can arise from various emission lines that are redshifted into the bandpass of the filter.
The main emission lines detected in our NB118 observations are as follows:

$\bullet$ \SIII\ $\lambda\lambda$9052,9532 at $z=0.31$ and 0.24, 

$\bullet$ \SII\ $\lambda\lambda$6717,6731 at $z=0.76$,

$\bullet$ \Ha\ at $z=0.80$,

$\bullet$ \OIII\ $\lambda\lambda$4959,5007 at $z=1.39$ and 1.36,

$\bullet$ \Hb\ at $z=1.44$,
and

$\bullet$ \OII\ $\lambda$3727 at $z=2.2$. 

To isolate the \Ha\ emitters in our sample, we use a combination of techniques. For \Pspecs\ [\Pspec] of the 3$\sigma$
[2.5$\sigma$] selected objects, spectroscopic data are available from our own targeted follow-up of New\Ha\ narrowband
excess sources (see below) with a small contribution from public spectroscopic datasets. For the remainder of the sample,
a classification scheme based on optical broadband colors is used.  The classification is empirically calibrated with
a combination of (1) the subset of NB118 excess emitters for which spectroscopy and optical broadband photometry
are available, and (2) galaxies in the COSMOS spectroscopic catalog whose redshifts would cause an
emission line to appear in the NB118 bandpass.   

Overall, we find that 48\% of our 3$\sigma$ narrowband excess sample (\NHas/\NNBs) can be identified as \Ha\ emitters.
The fraction is slightly lower (43\% or \NHa/\NNB) for the 2.5$\sigma$ sample.  Among our four fields, the fraction
ranges from 35\% to 65\%. This observed field-to-field variation is discussed in the context of predictions for
cosmic variance in Section \ref{sec:lf}. A summary of these statistics is given in Table~\ref{table2}.


\subsubsection{Spectroscopy of New\Ha\ Narrowband Excess Sources}\label{3.2.1}

We carried out spectroscopy of New\Ha\ narrowband excess sources between 2008--2009 with the Inamori Magellan Areal
Camera and Spectrograph \citep[IMACS;][]{dressler06} at the 6.5m Magellan I telescope.  IMACS enables multi-object
spectroscopy with slit masks over a 27\farcm4 diameter area (an excellent match to the field of view of NEWFIRM), and
has good sensitivity to $\sim$9500 \AA. These two characteristics make IMACS an ideal instrument for optical spectroscopic
follow-up of New\Ha\ NB118 excess sources, and in particular, \Ha\ emitters at $z=0.80$. Our observational
setup yields spectral coverage from 6300 \AA\ to 9600 \AA\ with 9 \AA\ resolution, and captures the complete ensemble
of strong rest-frame optical emission lines blueward of \Ha\ at $z=0.80$, from \OII\ $\lambda$3737 at 6720 \AA\ to
\OIII\ $\lambda$5007 at 9030 \AA. The spectra have a median integration time of $\sim$3.5 hr on average, reaching
5$\sigma$ sensitivities of $\sim2\times10^{-17}$ \fluxunit. IMACS spectroscopy has been obtained for
56\% [42\%] of our candidates, and the data are currently being used to probe the attenuation and metallicity properties
of the sample (I. Momcheva et al. 2011, in preparation). A more complete description of the New\Ha\ IMACS follow-up campaign is
provided in J. Lee et al. (2011, in preparation) and I. Momcheva et al. (2011, in preparation).

Additional spectroscopy is available from observations carried out by the SXDS (M. Akiyama, 2008, private communication)
and zCOSMOS \citep[DR2; ][]{lilly07} survey teams. The sample sizes of the spectroscopic catalogs are 11975 for COSMOS
and 4231 for SXDS, and among these, 2690 and 2386 fall in our survey regions. The zCOSMOS survey uses the VIMOS
spectrograph on the Very Large Telescope 8 m, and targets $I<22.5$ mag sources, as well as galaxies that are
color-selected to have $1.4<z<3.0$. The existing SXDS spectroscopic catalog is formed from the composite of data
from a number of individual observing programs targeting a range of SXDS sub-samples, using various instruments, and
achieving different depths. The mean and 1$\sigma$ dispersion of the magnitude distribution for the available SXDS
\zspec\ catalog are $\Rcf = 21.7$ and 2.0 mag, respectively. Both of the resultant SXDS and zCOSMOS
spectroscopic catalogs are dominated by objects that are bright relative to our NB118 excess emitters, and hence
the overlap with our sample is small (5\% or 58/1218).

\begin{figure} 
  \epsscale{1.2}
  \plotone{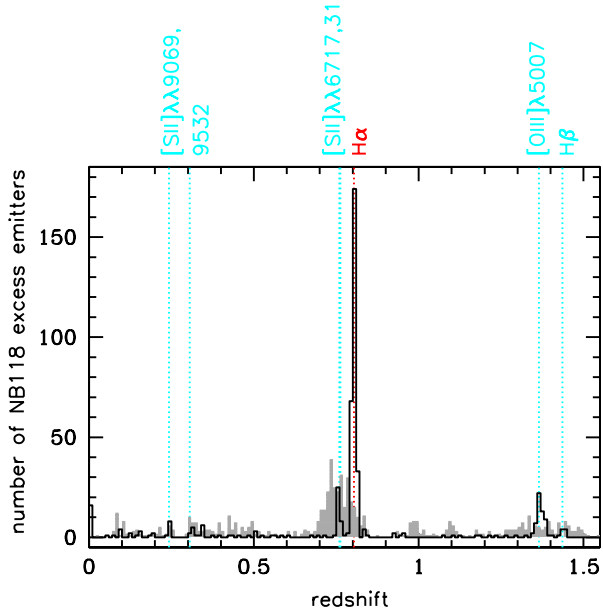}
  \vspace{-0.75cm}
  \caption{Redshift distributions for sources selected as NB118 excess emitters as determined from optical
    spectroscopy (black histogram) and from photometric redshifts (gray shaded histogram). The redshifts of the primary
    emission lines detected in our NB118 observations are labeled. (A color version of this figure is available in the online journal.)}
  \label{fig:specz}
\end{figure}

In total, 504 [566] sources in our NB118 excess sample have spectroscopic redshifts. The redshift
distribution for these galaxies is shown in Figure~\ref{fig:specz}. There are 253 [272] galaxies with
redshifts between 0.790 and 0.817, which places \Ha\ in the NB118 bandpass, and thus confirms them as \Ha\ emitters.
 
\subsubsection{Empirical Optical Color Classification}

\begin{figure*} 
  \epsscale{1.1}
  \plottwo{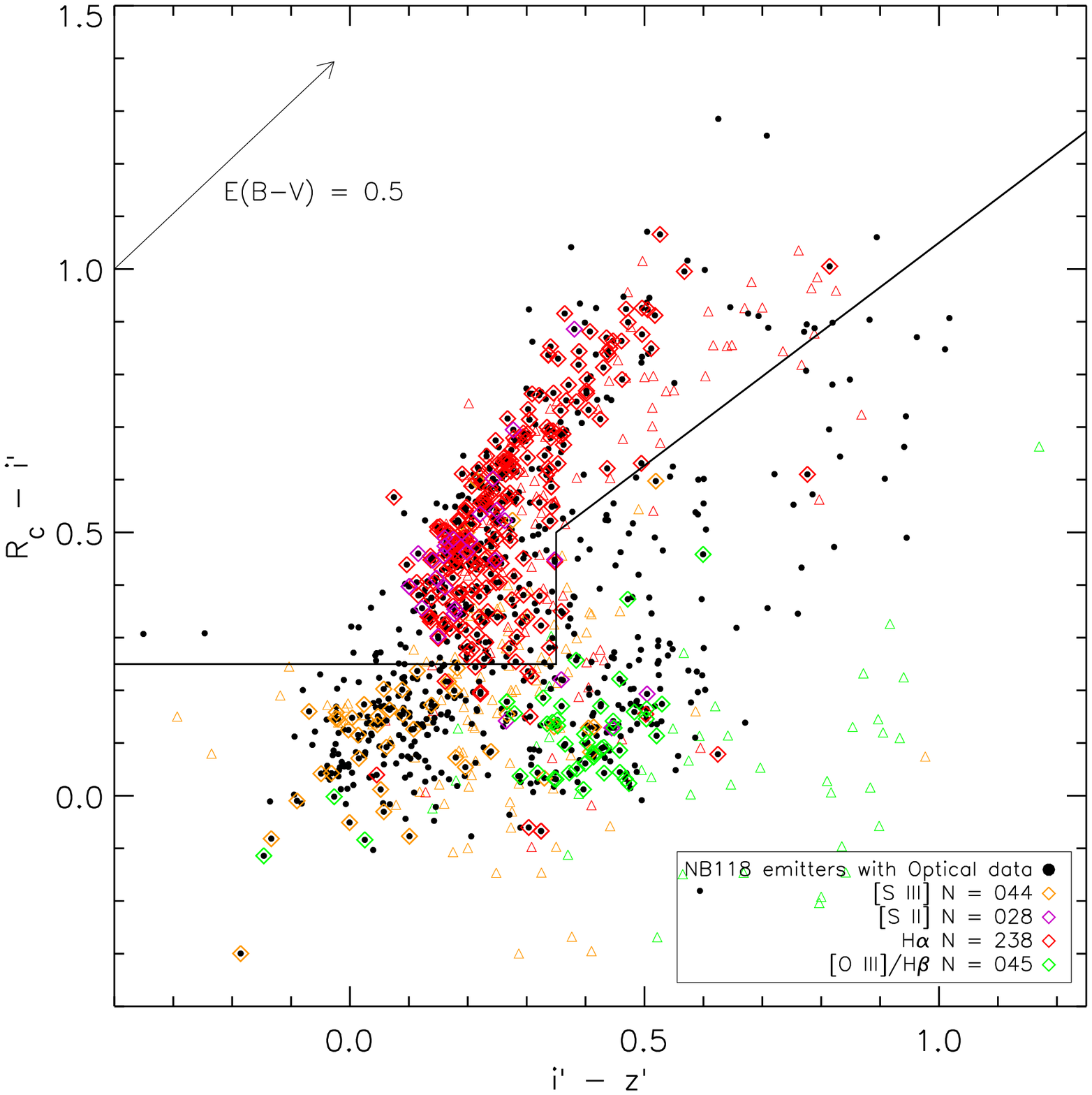}{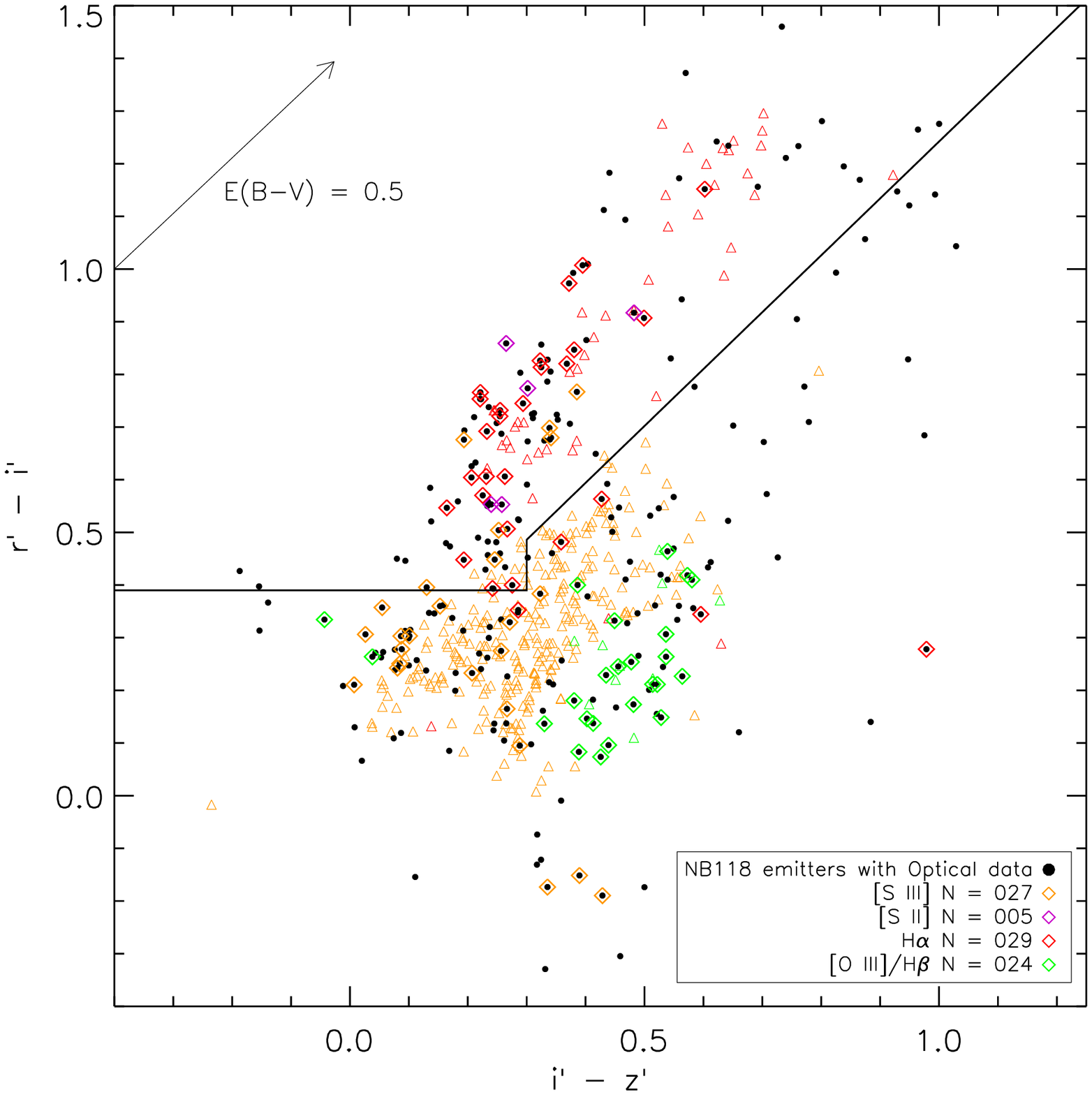}
  \caption{Left: \Rc\ $ - i\arcmin$ and $i\arcmin-z\arcmin$ colors of NB118 excess emitters in the three fields
    observed in the SXDS (black filled circles). Open diamonds are over-plotted on objects with spectroscopic redshifts,
    and are color-coded to indicate the emission line responsible for the narrowband excess, as described in the legend. 
    Right: analogous plot for the one field observed in COSMOS, but with $r^\prime$ instead of \Rc\ photometry. 
    To improve coverage of the diagram in this field, we also include galaxies from the zCOSMOS catalog whose redshifts
    would cause an emission line to appear in the NB118 bandpass (open triangles), but lie outside the area of our
    imaging. The black lines in both panels illustrate our adopted color criteria for the selection of \Ha\ emitters in
    the NB118 excess sample (see Equations~(\ref{eq:colorselectSXDS})--(\ref{eq:colorselectCOSMOS})).
    (A color version of this figure is available in the online journal.)}
  \label{fig:colorselect}
\end{figure*}

For the remaining 38\% [54\%] of the NB118 excess sources that do not have spectroscopic redshifts, a classification
scheme based on publically available Suprime-Cam broadband photometry is developed to separate \Ha\ emitters from other
sources in the sample. Our classification uses $R_{\rm C}\,i\arcmin\,z\arcmin$ in the SXDS fields, and
$r\arcmin\,i\arcmin\,z\arcmin$ in the COSMOS field. These sets of bandpasses are chosen because at $z=0.8$, the Balmer
and 4000 \AA\ breaks occur at $\sim$6500 \AA\ (in the $R$ band), so NB118 \Ha\ emitters will appear much redder in
$R - i\arcmin$ for a given $i\arcmin - z\arcmin$ than other narrowband excess sources (Figure~\ref{fig:colorselect}).

In the SXDS fields, we examine the distribution of NB118 excess sources that have spectroscopy from our follow-up IMACS
observations (Figure~\ref{fig:colorselect}, left panel), and identify \Ha\ emitters as those sources with
\begin{eqnarray}
  \label{eq:colorselectSXDS}
  \Rcf - i\arcmin \geq 0.25, \;\;\; \mbox{if} \; i\arcmin - z\arcmin \leq 0.35 \; \\
  \Rcf - i\arcmin \geq 0.846(i\arcmin - z\arcmin) + 0.204, \;\;\; \mbox{if} \; i\arcmin - z\arcmin > 0.35.  \;
\end{eqnarray}

In the COSMOS field, our IMACS spectroscopic dataset sparsely samples the color--color diagram. Thus, to improve coverage
of the diagram, we also include galaxies in the zCOSMOS spectroscopic catalog which lie outside the area of our
imaging, but whose redshifts would cause an emission line to appear in the NB118 bandpass (open triangles in
Figure~\ref{fig:colorselect}, right panel). Similar color criteria are adopted:
\begin{eqnarray}
  \Rcf - i\arcmin \geq 0.39, \;\;\; \mbox{if} \; i\arcmin - z\arcmin \leq 0.30 \; \\
  \Rcf - i\arcmin \geq 1.08(i\arcmin - z\arcmin) + 0.162, \;\;\; \mbox{if} \; i\arcmin - z\arcmin > 0.30.  \;
\label{eq:colorselectCOSMOS}
\end{eqnarray}
An additional 141 [250] NB118 excess sources lacking spectroscopic redshifts are identified as \Ha\
candidates with these criteria in both fields.

Of course, this color method is a blunt selection tool compared with the use of spectroscopic redshifts---it leads to
the inclusion of some interlopers, while missing true \Ha\ emitters that do not lie within the color selection region,
as evident from Figure~\ref{fig:colorselect}. In particular, the color criteria cannot distinguish between \Ha\ and
\SII\ emitters because the wavelengths of the emission lines are separated by $\sim$100 \AA. 

Fortunately, we can estimate the contamination and miss rates by examining the color classification of the NB118 excess
sources with spectroscopic redshifts. In the SXDS fields, 15\% of the excess sources with available spectroscopy that lie
in the \Ha\ selection region are incorrectly classified as \Ha\ emitters, and 5\% are confirmed \Ha\ emitters that
fall outside the color selection region.  The corresponding numbers for the COSMOS field are 29\% and 8\%. Applying these
rates to the number of \Ha\ emitters that are added to the \Ha\ sample via the color method, we find that 25 [45] sources
are potential interlopers and 8 [15] \Ha\ emitters could be missed. However, the {\it overall} contamination and miss
rates are far more limited with a high level of spectroscopic completeness of the sample: the estimated number of
interlopers and missed \Ha\ emitters are only 6\% [2\%] and 9\% [3\%] of the total \Ha\ sample.


\section{Calculation of \Ha\ Luminosities}\label{2.3}

\subsection{Observed Line Fluxes, EWs and Luminosities}
\begin{figure} 
  \epsscale{1.1}
  \plottwo{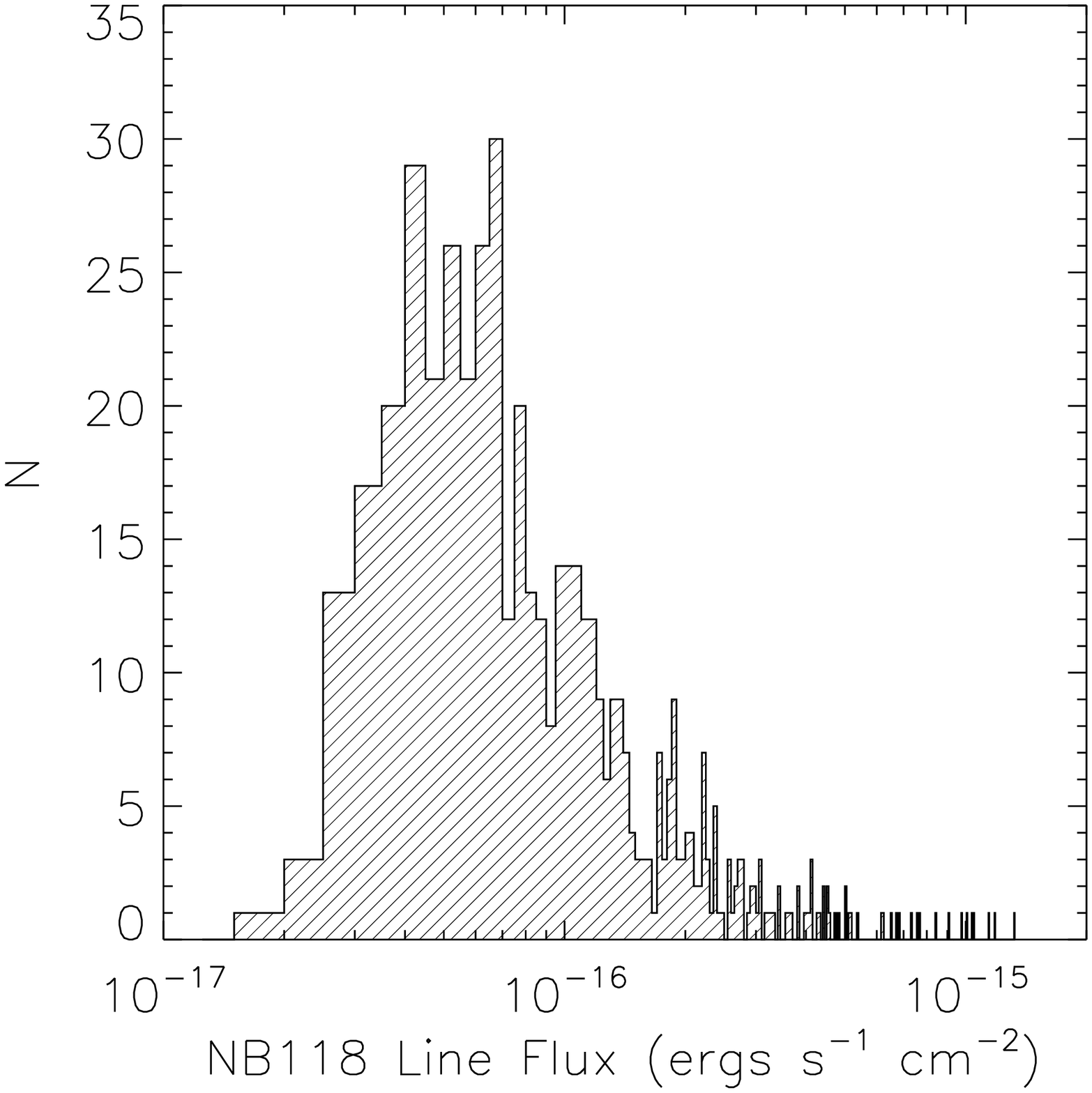}{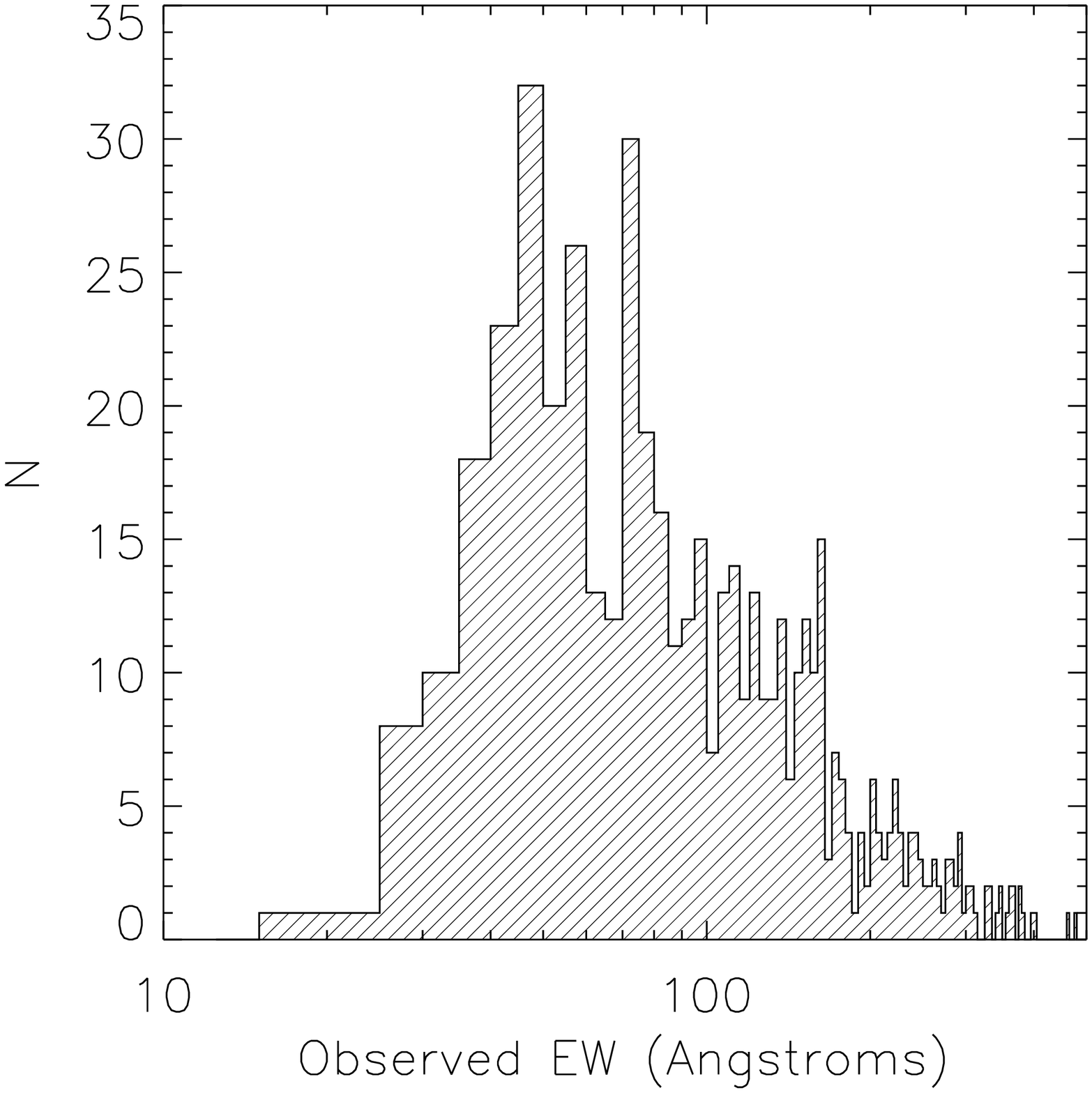}
  \caption{Distribution of the observed \Ha\ + \NII\ emission-line fluxes (left) and EWs
    (right) for all identified \Ha\ emitters, computed from photometry measured in the
    larger choice of apertures, as discussed in Section \ref{sec:NB118select}.
    The EWs are a factor of $1+z$ larger than the rest-frame EWs.}
  \label{fig:ewfluxdist}
\end{figure}

Emission-line fluxes and EWs for the $z=0.8$ \Ha\ galaxy sample are calculated as follows. Flux densities
(\fluxunit\ \AA$^{-1}$) may be written as $f_{\rm NB}=f_C+F_L/\Delta$NB and $f_{J}=f_C + F_L/\Delta J$,
where $f_C$ is the continuum flux density, $F_L$ is the emission-line flux (\fluxunit), and $\Delta$NB and
$\Delta J$ are the full width at half maximum of the NB118 (110 \AA) and $J$ (1786 \AA) filters. Therefore, the
emission-line flux and the continuum flux density are computed as
\begin{eqnarray}
  F_{L} = &\Delta{\rm NB} \frac{f_{\rm NB} - f_{J}}{1-(\Delta{\rm NB}/\Delta J)},~{\rm and}\\
  f_C = &\frac{f_{J} - f_{\rm NB}(\Delta{\rm NB}/\Delta J)}
  {1-(\Delta{\rm NB}/\Delta J)}.
\end{eqnarray}
By definition, the observed EW, which is a factor of $1+z$ larger than the rest-frame EW, is the
ratio of the emission-line flux to the continuum flux density, and thus
\begin{equation}
  {\rm EW}_{\rm obs} \equiv \frac{F_L}{f_C} = \Delta{\rm NB}\frac{f_{\rm NB} - f_{J}}{f_{J} - f_{\rm NB}(\Delta{\rm NB}/\Delta J)}.
\end{equation}
With this equation, the minimum $\Delta(J-{\rm NB118}) = 0.2$ mag excess corresponds to an observed EW of 40 \AA.
To compute luminosities, we adopt a distance of 5041 $h_{70}^{-1}$ Mpc, which corresponds to $z=0.803$, the median
redshift of our spectroscopically confirmed \Ha\ emitters. Distributions of the observed emission-line fluxes and EWs
for our sample of \Ha\ emitters are shown in Figure~\ref{fig:ewfluxdist}.

\subsection{Derived \Ha\ Luminosities}

To determine the intrinsic \Ha\ luminosity from the observed line luminosity, corrections must be applied for
contamination of the flux by the adjacent \NII\ lines, and for attenuation by dust internal to the galaxy. The
corrections that we adopt are based on standard, empirically calibrated relationships, which describe average
corrections for ensemble populations. They should yield luminosities appropriate for computing averaged/integrated
quantities such as the SFR volume density. However, it should be noted that the corrections will not be accurate for
individual galaxies, since the scatter in the relationships is large, as described below.

{\it \NII\ contamination.} The NB118 bandpass is wide enough to include flux from the
\NII\ $\lambda\lambda$6548,6583\footnote{Hereafter, ``\NII'' refers to both nitrogen nebular emission lines.} emission lines
for narrowband excess selected \Ha\ emitters.

In the local universe, integrated spectroscopic surveys have found that the \Ha/\NII\ flux ratio is 2.3 for typical
$\Lstar$ galaxies \citep{kennicutt92,gallego97}, and past narrowband surveys have often adopted a fixed correction 
of 2.3 for all selected \Ha\ emitters. The flux ratio, however, has been shown to increase with larger emission-line
EW \citepalias[see, e.g.,][]{villar08} and with decreasing $B$-band luminosities \citep{kennicutt08,lee09}.
Such correlations are likely a consequence of the mass--metallicity relationship \citep[e.g.,][]{lee04,tremonti04}.
In a recent deep optical narrowband survey with Subaru Suprime-Cam, \cite{ly07} adopted 4.66 based on optical
spectroscopic follow-up of their emitters. The sample of \cite{ly07} consists of galaxies with fainter luminosities,
hence are more metal-poor and would have a higher \Ha/\NII\ flux ratio on average.
 
For this study, we follow \citetalias{villar08} and \citetalias{sobral09} in adopting an EW-dependent \Ha/\NII\ flux
ratio to facilitate comparisons of results. The EW-dependent correction was constructed from thousands of $z\sim0.1$
star-forming galaxies from the Sloan Digital Sky Survey fourth data release with which \citetalias{villar08} determined
the mean relationship between the rest-frame EW of \Ha\ + \NII\ $\lambda$6583 and the \Ha/\NII\ flux ratio. 
The correlation exhibits a scatter of $\sim$0.2 dex, which implies that estimates of the \Ha/\NII\ ratio for
individual sources will only be accurate to $\sim$50\%. Such errors will average out in the calculation of
integrated quantities however, and are adequate for our purposes here.

Also, our correction assumes that the \Ha/\NII\ relation does not evolve with redshift. This is a valid
assumption since the evolution of metallicity between $z=0.07$ and $z=0.7$ has been found to be no more than 0.1 dex
\citep[see, e.g.,][and references therein]{mannucci09}. Near-infrared multi-object spectroscopy will be
needed to examine the \Ha/\NII\ ratio for NB selected galaxies on a case-by-case basis. Assuming the \citetalias{villar08}
\NII\ correction, the \Ha/\NII\ ratio varies from 1.85 to 10.0 with a median (average) of 2.50 (2.81) for our
population of \Ha\ emitting galaxies.

{\it Dust attenuation.} To correct for dust attenuation, we adopt a luminosity-dependent extinction relation
following \cite{hopkins01}:
\begin{eqnarray}
  \nonumber
  \log{\left[{\rm SFR}_{{\rm obs}}({\rm H}\alpha)\right]}&=\log{\left[{\rm SFR}_{\rm int}({\rm H}\alpha)\right]}-2.360\\
  &\times\log{\left[\frac{0.797\log{\left[{\rm SFR}_{\rm int}({\rm H}\alpha)\right]}+3.786}{2.86}\right]}.
  \label{ext}
\end{eqnarray}
\cite{hopkins01} showed that attenuations computed from this equation show a 0.2 dex scatter relative to those
based on Balmer decrement measurements for individual galaxies.
Much like the \NII\ correction, these dust extinction corrections are more
reliable when considering multiple sources of a given luminosity/SFR. The \cite{hopkins01} relation
was also adopted in other studies (e.g., \citealt{ly07}; \citetalias{villar08}; \citealt{dale10}).
For our sample, this correction (A[\Ha]) ranges from 0.66 to 1.87 mag with a median and average of
1.19 and 1.21 mag, respectively.

One piece of evidence that supports the \cite{hopkins01} relation is the work of \cite{garn10}, which has
looked at the UV, \Ha, and mid-infrared fluxes for a sample of narrow-band selected \Ha\ emitters to
determine dust attenuation. They find a correlation between the observed \Ha\ luminosity and dust
attenuation that is similar to \cite{hopkins01}. This multi-wavelength comparison will be
conducted for our sample of \Ha\ emitters. We will also investigate dust attenuation determined from
Balmer decrements, which are obtained from the combination of our NB118 measurements and
our IMACS follow-up spectroscopy (I. Momcheva et al. 2011, in preparation). Preliminary results do indicate
that the attenuations based on Balmer decrements for the NB118-selected \Ha\ emitters are
consistent with the \cite{hopkins01} correction.

\section{Completeness of the Survey}\label{3}

\subsection{Monte Carlo Simulations}\label{3.1}
{\it Motivation.}
To account for the detection limits and photometric selection of the New\Ha\ survey, we need to estimate the completeness
fraction as a function of luminosity, \C. The greatest uncertainty for such a task is the {\it intrinsic} EW
distribution, that is, the inherent distribution of the population that we are able to seek. Since the narrowband and
broadband imaging sensitivities of the survey are known, the completeness is determined for a source
of a given brightness with a $J-{\rm NB118}$ excess color (i.e., an EW). However, sources of a given emission-line
luminosity can be faint with large EWs or bright with low EWs. Thus, the adopted EW distribution affects the estimated
completeness for individual emission-line luminosity bins. For this reason, we conduct Monte Carlo realizations of our
data, and follow a ``maximum likelihood'' approach to determine the completeness of New\Ha.

{\it Technique/approach.}
The methodology intends to simultaneously reproduce the shape of the
observed\footnote{Throughout this section, we use ``observed'' to denote that the EW distributions
  are a factor of $1+z$ larger than the rest-frame EW distributions.}
\Ha+\NII\ EW distribution and the observed \Ha\ LF. We begin with a grid of models that adopt certain EW distributions.
With these distributions, we generate artificial emission-line galaxies. We add noise to these galaxies, and use their
measured magnitudes and colors to determine if these mock galaxies satisfy the NB118 excess selection. We construct
the observed EW distribution and \Ha\ LF from the sample of mock NB118 excess emitters to compare to the respective
distributions from New\Ha. We determine a best model and define the completeness, \C, as the ratio of the number of
mock NB118 excess emitters to the total number of generated artificial galaxies:
\begin{equation}
  \kappa(L) = {\rm N}(L)_{\rm obs} / {\rm N}(L)_{\rm tot}.
\end{equation}

\noindent{\it Assumptions.}
For the prior distributions, we make two assumptions. First, for the \Ha\ LF, a proper normalization of bright-to-faint
sources is necessary, otherwise we will generate relatively too many bright sources, and a decline in the simulated
LF will not been seen and cannot be compare to the observed \Ha\ LF. To constrain the ratio of bright-to-faint
sources, we examine the galaxy number counts (N[$J$]) in the New\Ha\ fields and find that
$\log{[{\rm N}(J)]} \propto 0.344 \times J$. This relation implies that the number of galaxies at magnitude
$J\arcmin+1$ is a factor of 2.2 more numerous than those at $J\arcmin$.

Second, each plausible model assumes a log-normal EW distribution described by a mean $\mumod$ and a sigma $\sigmamod$.
We adopt such a distribution, since it is fairly similar to the observed New\Ha\ distributions and the distributions
seen locally \citep{lee07}. Our models span $\mumod = 1.15$--1.55 and $\sigmamod = 0.15$--0.40 both in increments of
0.05 dex for a total of 54 EW distributions. Here, $\mumod$ is the logarithm of the {\it rest-frame} EW, and we later
include a factor of 1+$z$ for the ``observed'' magnitudes, colors, fluxes, and EWs.

\begin{figure}[htc] 
  \epsscale{1.1}
  \plotone{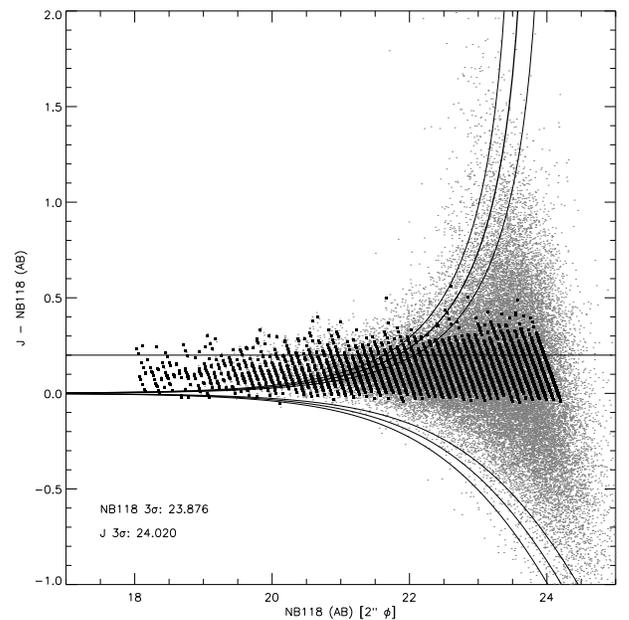}
  \caption{Color--magnitude diagram ($\Delta(J-{\rm NB118})$ vs. NB118) illustrating the selection of {\it mock} NB118
    \Ha\ excess emitters in the SXDS-W for the NE detector. We show modeled galaxies as black filled squares, and the
    gray points are the repeated measurements of these galaxies with photometric noise included. This figure represents
    2,004,800 mock galaxies with every fiftieth point shown for lower resolution. The horizontal line is the minimum
    0.2 mag excess that we adopt, and the three black curves show the 2$\sigma$, 2.5$\sigma$, and 3$\sigma$ $\Delta(J-{\rm NB118})$ color
    excesses. The model assumes $\mumod=1.15$ and $\sigmamod=0.15$.}
  \label{mock_scatter}
\end{figure}

{\it Implementation.}
For each of the 54 models, we generate $\sim$40,000 mock galaxies with $J$-band (continuum) magnitudes chosen to
follow the above relation between $\log{[{\rm N}(J)]}$ and $J$. We assume that these galaxies are spatially unresolved
and follow the FWHM of seeing for each mosaic. For the EWs, we randomly draw from a log-normal distribution given
by $\mumod$ and $\sigmamod$. With these two information, we ``fill'' the color--magnitude diagram ($\Delta(J-{\rm NB118})$
versus NB118; see Figure~\ref{mock_scatter}).

We add observational uncertainties to the $J$ and NB118 magnitudes as follows. For each mock galaxy, we have a randomly
chosen position in the mosaics, and generate 200 realizations based on the sensitivity at that position. The
sensitivity is governed by the exposure maps for both the NB118 and $J$-band mosaics, thus accounting for detector
and pixel-to-pixel variations. We compute the NB118 or $J$-band 3$\sigma$ limiting magnitudes, $m_{\rm lim}(x,y)$,
signal-to-noise ratios, S/N$(x,y)$, and 1$\sigma$ uncertainties, $\Delta m(x,y)$, as
\begin{eqnarray}
  m_{\rm lim}(x,y) = &~\langle m_{\rm lim}\rangle + 2.5\log{\left(\sqrt{t(x,y)/t_0}\right)},\\
  {\rm S/N}(x,y)  = &~3 \times 10^{-0.4[m - m_{\rm lim}(x,y)]},~{\rm and}\\
  \Delta m(x,y)   = &~2.5 \log{\left[1+\frac{1}{{\rm S/N}(x,y)}\right]},
\end{eqnarray}
where $\langle m_{\rm lim}\rangle$ and $t_0$ are the individual detector's median limiting magnitudes and exposure times
(see Table~\ref{table1}), respectively, and $m$ is the NB118 or $J$ magnitude. For the $\Delta(J-{\rm NB118})$ colors, the
errors are added in quadrature: $\sqrt{\Delta m_{\rm NB118}(x,y)^2+\Delta m_J(x,y)^2}$. The magnitudes for the 200
realizations follow a Gaussian distribution with a mean of $m$ and 1$\sigma$ of $\Delta m(x,y)$. 

We identify mock galaxies that are NB118 excess emitters using the same methods described in Section \ref{sec:NB118select}.
For the smaller apertures (2\arcsec\ or 2\farcs5), we assume that the flux enclosed is between 81\% and 86\% of the
total flux within the larger apertures (3\arcsec\ or 4\arcsec). We determine these values by considering a Gaussian
distribution with a FWHM equal to the seeing size for each NEWFIRM pointing. Note that we adopt limiting magnitudes
appropriate for the size of the measurement aperture.

\begin{figure} 
  \epsscale{1.1}
  \plotone{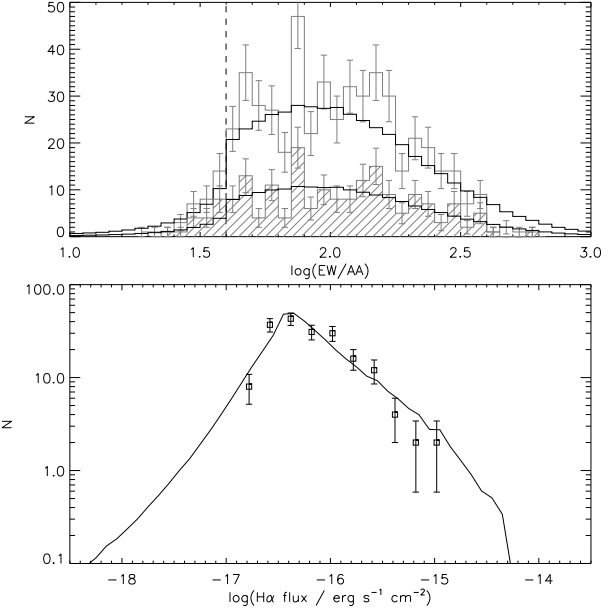}
  \caption{Comparisons between the modeled predictions and observations of the EW distribution (top) and the
    \Ha\ galaxy number counts (bottom) for the SXDS-W field. The shaded and unshaded histograms are for the
    SXDS-W and all four New\Ha\ fields, respectively. The dashed line in the top panel is the minimum adopted
    $J-{\rm NB118}$ excess of 0.2 mag or an observed EW of 40 \AA. The model (black curves) assumes $\mumod=1.35$ and
    $\sigmamod=0.40$, and is scaled to match the observed distributions. This model is able to produce the shape of
    the observed EW distribution and the \Ha\ number counts.}
  \label{mock_compare}
\end{figure}

{\it Comparisons with observations.}
For each model, we construct the observed EW distribution and \Ha\ LF from the mock NB118 excess emitters. Since we
generate many mock galaxies, we normalize each distribution to best match the observed New\Ha\ distributions. We
illustrate an example of these model--observation comparisons in Figure~\ref{mock_compare} for the SXDS-W field.
We find the best fit by minimizing $\chi^2$ through a comparison of the mock and the New\Ha\ EW distributions and
\Ha\ LFs. We show in Figure~\ref{mock_compare} that the best-fit model is able to reproduce both the EW distribution
and the \Ha\ LF. We find that observed EW distribution provides a better constraint on the best-fit model (versus
the \Ha\ LF). That is, multiple EW models can sufficiently explain the \Ha\ LF. We find the model with $\mumod=1.35$
and $\sigmamod=0.40$ best matches the observed distributions.

\begin{figure}[htc] 
  \epsscale{1.1}
  \plotone{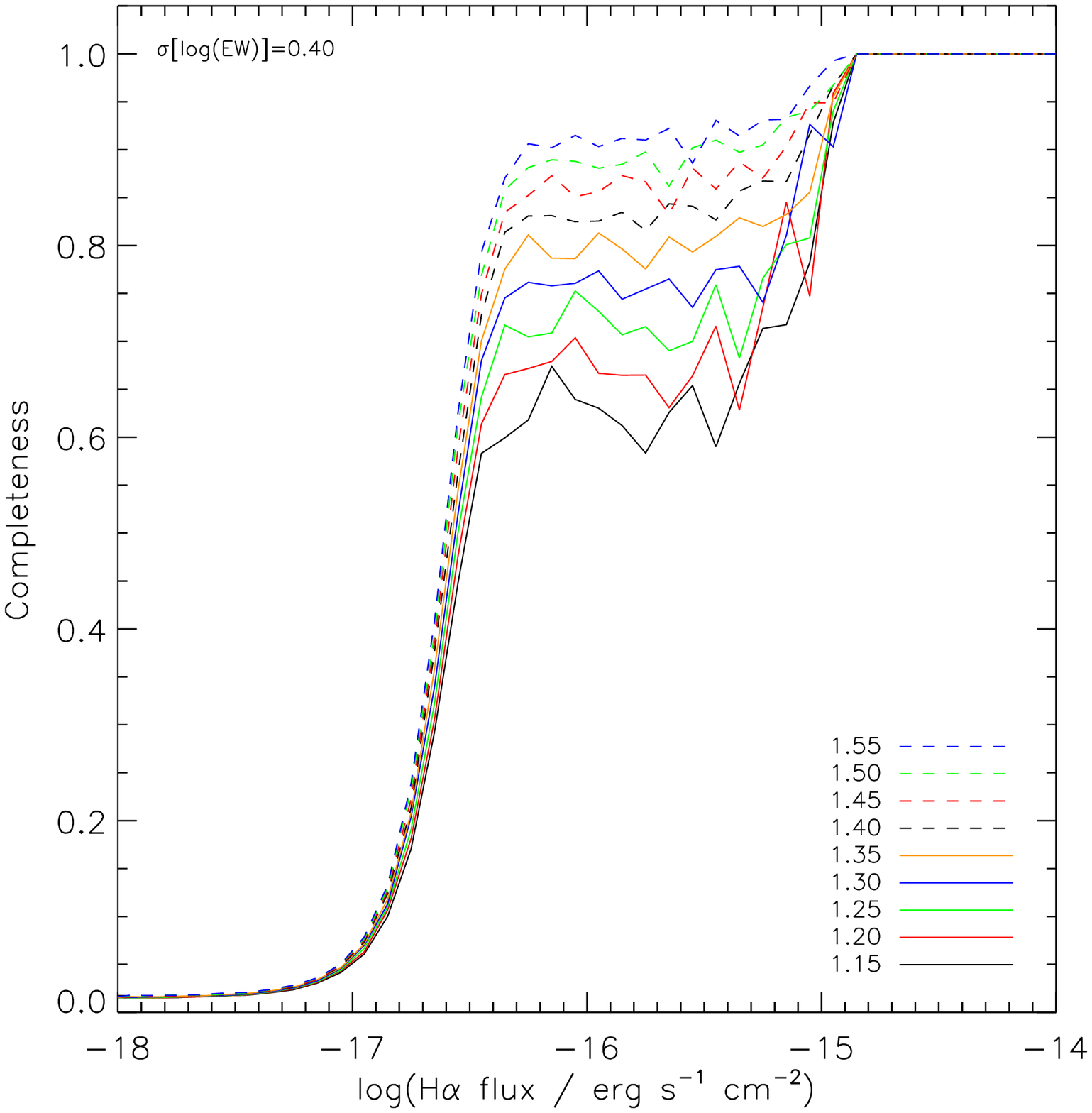}
  \caption{Completeness correction derived from Monte Carlo simulations of our SXDS-W data for models with
    $\sigmamod = 0.40$. The \Ha\ fluxes have not been corrected for any dust extinction. The color and line style
    conventions are for different adopted $\mumod$, as indicated in the lower right. This figure illustrates that as
    a larger $\mumod$ is adopted, the completeness increases for bright emission-line galaxies. As expected, the
    adopted minimum $J-{\rm NB118}$ color excess of 0.2 mag is unable to identify galaxies that are very bright with
    low EWs. (A color version of this figure is available in the online journal.)}
  \label{SXDSW_comp}
\end{figure}

\begin{figure*}[htc] 
  \epsscale{1.1}
  \plottwo{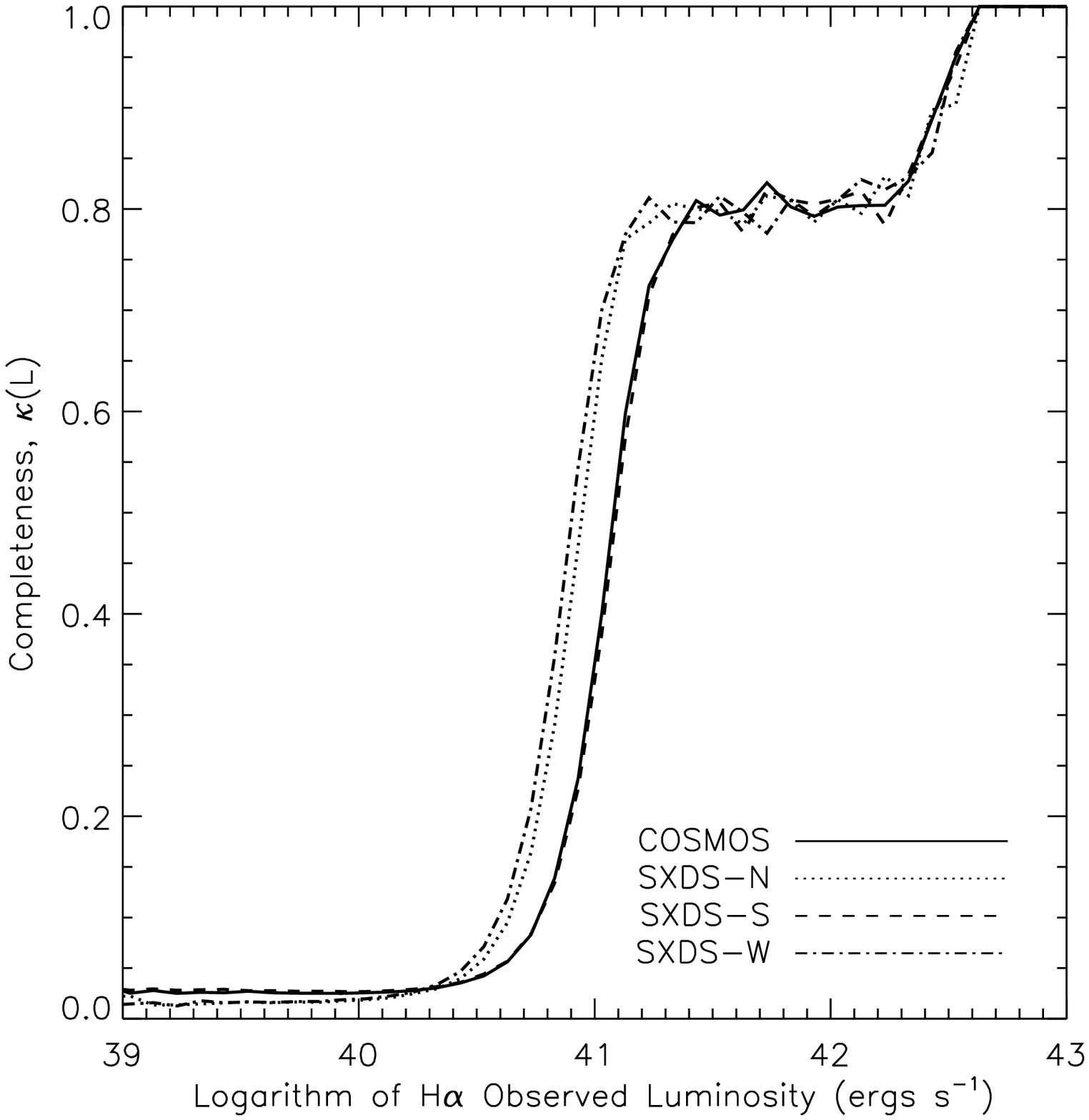}{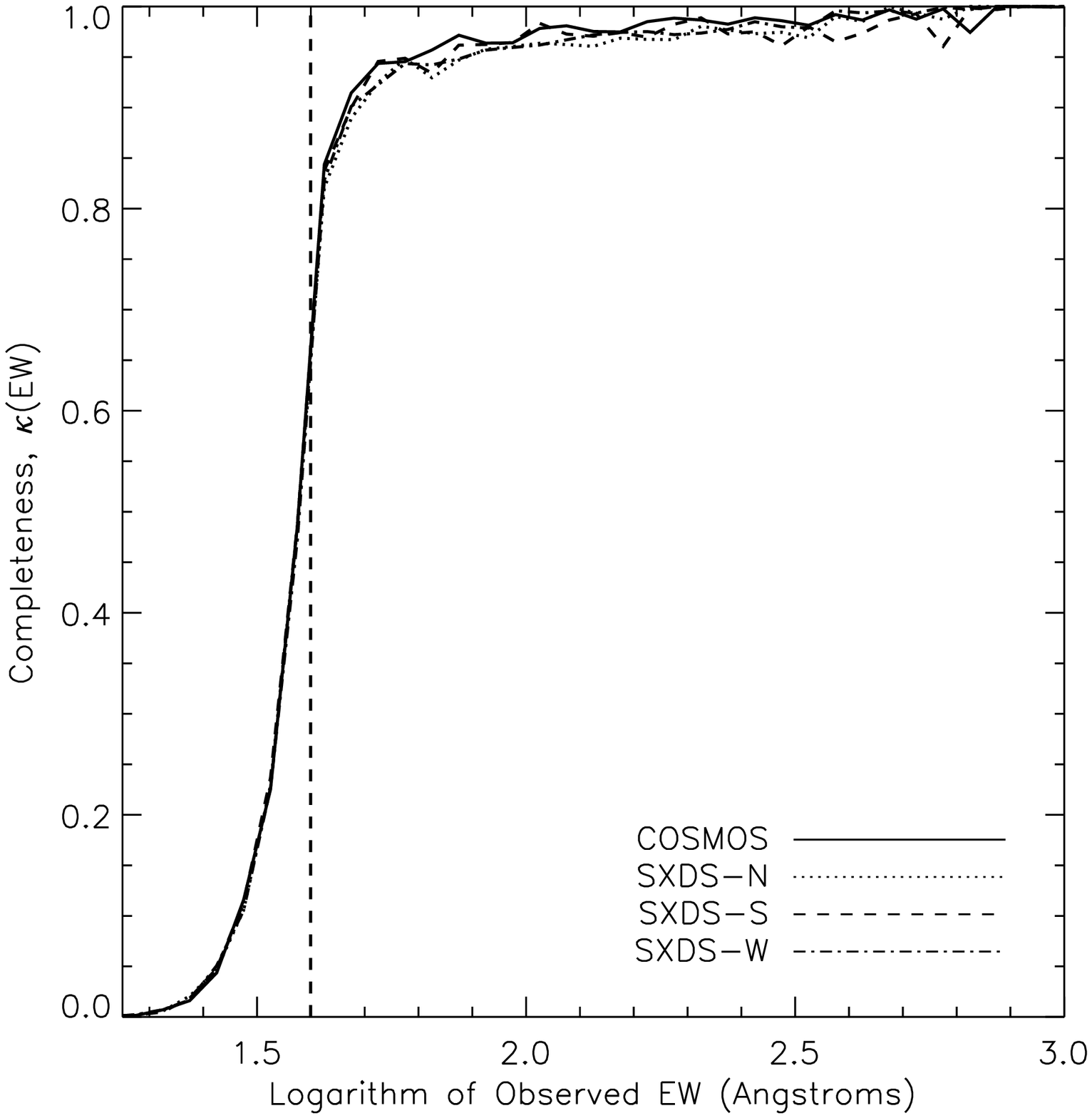}
  \caption{Completeness corrections as a function of luminosity ($\kappa(L)$; left) and
    EW ($\kappa({\rm EW})$; right) derived from Monte Carlo simulations of our data. The line types denote different
    fields: COSMOS (solid), SXDS-N (dotted), SXDS-S (dashed), and SXDS-W (dot-dashed). The best model
    ($\mumod=1.35$ and $\sigmamod=0.40$) is used. The higher completeness for SXDS-W and SXDS-N is due to higher
    spatial resolution and sensitivity. The vertical dashed line in the right panel is for the minimum observed
    EW limit of 40 \AA. For $\kappa({\rm EW})$, we consider mock galaxies with an \Ha\ emission-line flux
    above the $2.5\sigma$ flux limit of $\approx1.2\times10^{-17}$ \fluxunit, which is $\approx1.8\times10^{-17}$
    \fluxunit\ when the adjacent \NII\ emission is included (i.e., the total NB118 emission-line flux).}
  \label{EW_comp}
\end{figure*}

We calculate the completeness as the ratio of the number of mock galaxies that meets the NB118 excess selection
criteria to {\it all} mock galaxies ($N \approx 200 \times 40,000$) as a function of $L({\rm H}\alpha)$. We illustrate
in Figure~\ref{SXDSW_comp} how the completeness differs by adopting different $\mumod$ values. This comparison shows
that we miss bright galaxies with low EWs, which is not surprising given the minimum observed EW of 40 \AA. This figure
also reveals that the completeness can vary between 60\% and 90\% for observed fluxes of
$5\times10^{-17}$--$6\times10^{-16}$ \fluxunit\ or roughly 0.1--1 $\Lstar$ 
(see Section \ref{sec:lf}), and emphasizes that proper choice in the prior EW distribution is necessary for an accurate
determination of the LF.
We also show the completeness as a function of \Ha\ luminosity and observed EW for each of the fields assuming the
best-fitting model with $\mumod = 1.35$ and $\sigmamod = 0.40$ in Figure~\ref{EW_comp}.

{\it Discussion of simplifying assumptions.}
In the above calculations, we made a couple of simplifications to allow the Monte Carlo simulations to be significantly
less computationally intensive. First, we assume that the galaxies are unresolved sources. Second, instead of detecting
sources on the images, we estimate the measured magnitudes and fluxes where the photometric uncertainties are based
on pixel-to-pixel and detector-dependent sensitivity.
Third, it is important to note that our current simulations assume that the EW distribution is
the same at all continuum magnitudes. It would be more correct to allow the EW distribution to
vary as a function of continuum magnitude, since it has been shown that more luminous
star-forming galaxies tend to have lower emission-line EWs than those at lower luminosity
\citep[e.g.,][]{lee07}. However, this adds another degree of freedom to our models, and
significantly increases the computational time required to complete the simulations. Future
work will include greater complexity in the modeling of the EW distribution, and will examine
the impact of the current assumptions on the resultant completeness corrections.
These simplifications allow for a few orders of magnitude faster
production of completeness results rather than the approach of adding sources directly to the images, and is able to
reproduce the distributions of scatter in Figures~\ref{fig:zJ} and \ref{fig:colormag}
(e.g., see Figure~\ref{mock_scatter}).

Prior to these MC simulations (for this discussion, we refer to the above approach as ``Method II''), we generated
a simulation where artificial extended sources were directly added to the mosaics and detected. In this
simulation (referred to as ``Method I''), we adopted a log-normal Gaussian EW distribution with $\mumod=1.52$ and
$\sigmamod=0.16$, motivated by the results in \cite{lee07}. 
We performed Method I by adding 1000 artificial galaxies to each NEWFIRM mosaic, and compared the number of artificial
NB118 excess emitting galaxies to the number detected. We created the galaxies using IRAF/{\tt mkobjects} with physical
parameters similar to those found in local galaxies. The parameters include luminosities, \Ha\ EWs, semimajor to
semiminor axial ratios (between 0.15 and 1.0), and the \Ha\ disk scale length \citep[3.6 kpc;][]{dale99}.
After extracting sources in the same manner described in Section \ref{sec3}, we calculated the completeness. We repeated
each simulation 30 times using different seed numbers to increase the statistical accuracy of the results
($N\sim30,000$ per NEWFIRM pointing).
The similarities in the shape of the completeness curves in Methods I and II suggest that the simplifications
we made do not have a significant effect on the estimated completeness. There are certain differences such that the
comparison is not completely ``apples-to-apples''; however, good agreement suggests that proceeding with Method II
is satisfactory. The greatest benefit of this simulation is the determination of a model that best matches two
different sets of measurements.

Finally, we assume a power-law distribution for N($J$) to normalize the number of bright and faint
sources. Ideally, the distribution should be constructed from galaxies at $z \sim 0.8$.
An investigation with the COSMOS photo-$z$ sample (after the simulation was completed) finds that the
faint-end slope of the number counts is approximately 0.25, which is to be compared to the adopted 0.34.
Also, an exponential decline exists at bright continuum luminosities. This will impact our completeness
estimates for the brightest emitters. Future work will address and examine the impact that these
differences have on the completeness reported here.

\subsection{Effective Volume}\label{3.2}
The volume that the survey is capable of probing is determined by the shape and width of the NB filter.
The ideal filter will have a perfect top-hat profile, and will survey the same co-moving volume at all
emission-line fluxes.
However, with real filters 
a weak emission line can either result from the line falling in the wings of the NB filter profile or
from an intrinsically weak line located near the filter center.
The net result is that a weak emission line is more likely
to be detected near filter center, so a non-square filter reduces the effective volume at the faint end \citep{ly07}.

We can quantify how much the filter profile deviates from being a perfect top-hat by comparing the area underneath
the profile with that of a rectangle with width equal to the FWHM, and height equal to the maximum transmission
of the actual filter.  The area enclosed by the profile is just 9\% smaller, so the shape of the bandpass is close
to the ideal.

\begin{figure}[htc] 
  \epsscale{1.1}
  \plotone{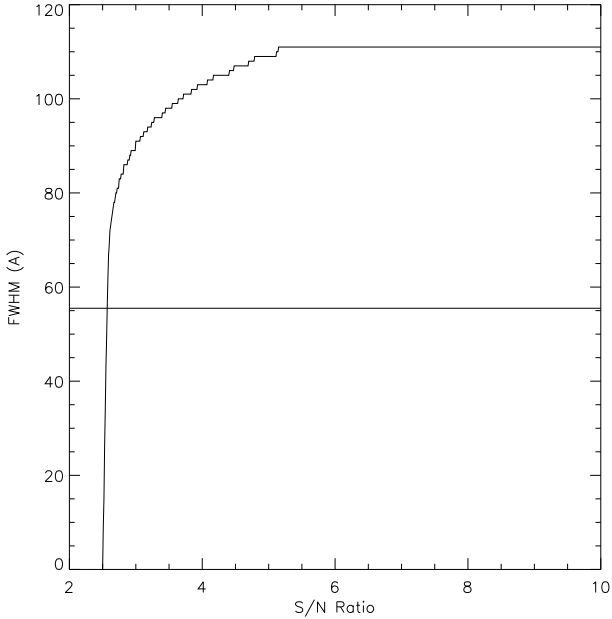}
  \caption{Effective surveyed volume. The FWHM (in wavelength) of the surveyed volume versus S/N ratio. The
    horizontal line corresponds to 50\% of the maximum effective volume
    ($V_{\rm eff}/\Omega = 1.11\times10^5$ Mpc$^{3}$ deg$^{-2}$).}
  \label{Veff}
\end{figure}

To determine the effective volume as a function of emission-line flux, we calculate the range in wavelengths such that
an emission line is considered detectable within the NB filter. This emission line has an intrinsic S/N. We then place
it at different wavelengths to determine what the degradation in the S/N will be due to lower throughput. We define
an emission line to be undetected below 2.5$\sigma$, and this criterion yields the minimum and maximum NB118
wavelengths (hence redshift) that is observable for a particular S/N. The effective comoving volume per unit steradian
would then be
\begin{eqnarray}
  \frac{V}{d\Omega} & = & \int_{z_1}^{z_2} dz\frac{dV}{dz d\Omega}, {\rm where}\\
  \frac{dV}{dz d\Omega} & = & \frac{c}{H_0}\frac{D_M^2}{E(z)}, \\
  E(z) & \equiv & \sqrt{\Omega_M(1+z)^3 + \Omega_{\Lambda}}, {\rm and}\\
  D_M & = & \frac{c}{H_0}\int_0^z \frac{dz\arcmin}{E(z\arcmin)}.
\end{eqnarray}
Here, $z_1$ and $z_2$ refer to the minimum and maximum redshifts that the emission line is detectable. Note that these
equations assume a flat universe. The maximum surveyed volume ($V_{\rm eff}/\Omega = 1.11\times10^5$ Mpc$^{3}$
deg$^{-2}$ or $\Delta\lambda = 110$ \AA) is observable for ${\rm S/N} \geq 5.1$ and decreases to
$V_{\rm eff}/\Omega = 5.54\times10^4$ Mpc$^{3}$ deg$^{-2}$ ($\Delta\lambda = 55$ \AA) at ${\rm S/N} = 2.57$. This is
illustrated in Figure~\ref{Veff}.

\section{Results}\label{4}
\input tab3_arxiv.tex

\subsection{Luminosity Function and Star Formation Rate Densities}\label{sec:lf}
\begin{figure*}[htc] 
  \epsscale{1.1}
  \plottwo{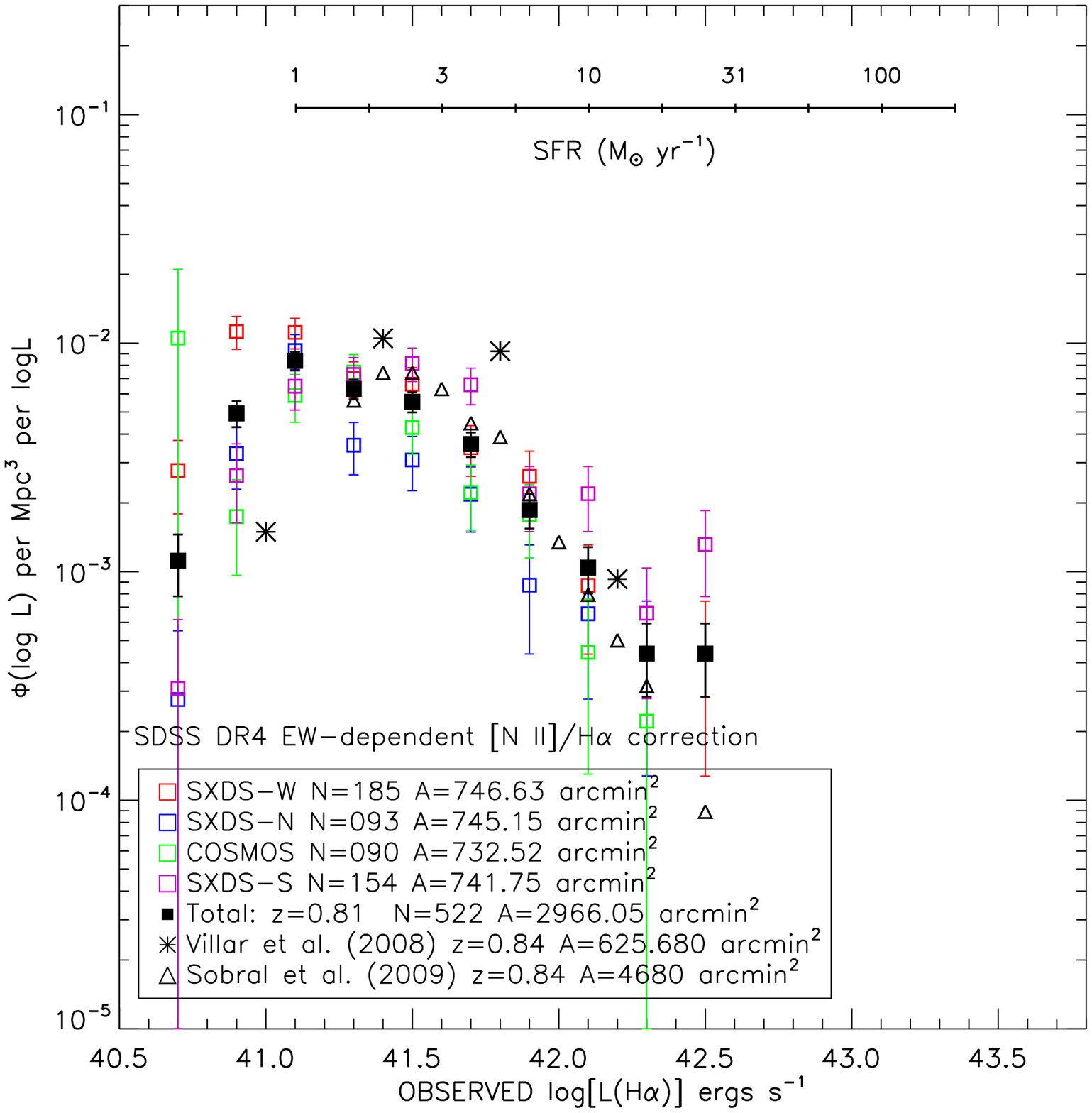}{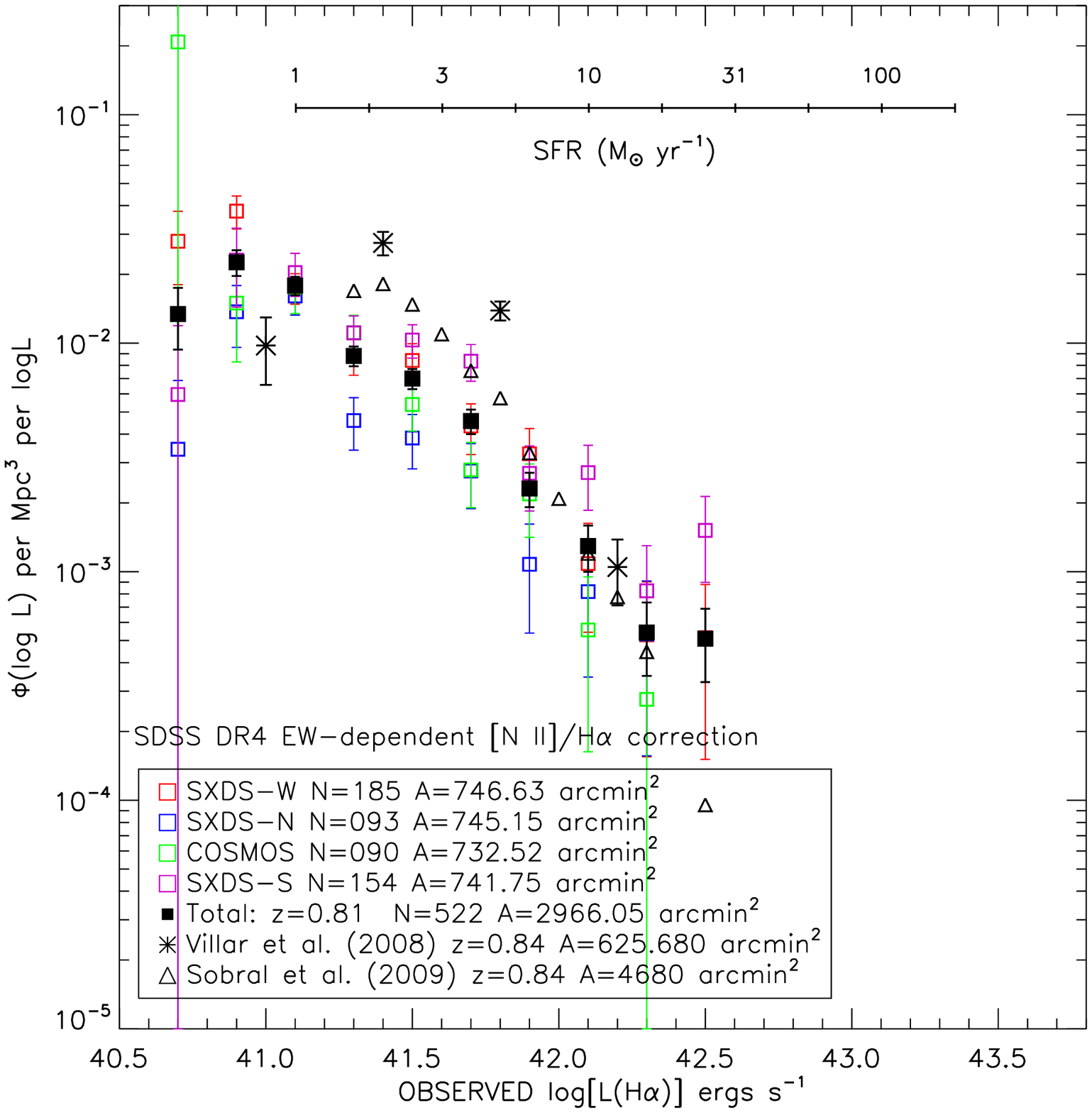}
  \caption{Observed (i.e., no dust attenuation corrections have been applied) \Ha\ LF before (left)
    and after (right) applying completeness correction. Corrections have also been made for \NII\ contamination.
    The combined measurements from the four New\Ha\ fields are shown as filled squares, where as open squares
    represent the individual pointings. \citetalias{villar08} and \citetalias{sobral09} measurements are shown as
    asterisks and triangles, respectively. The axis range is kept the same as other plots of the LF for comparison
    purposes. All LFs have selected sources down to 2.5$\sigma$ significance. Conversion between \Ha\
    luminosities and SFRs is based on the \cite{kennicutt98} relation.
    (A color version of this figure is available in the online journal.)}
  \label{HaLF}
\end{figure*}

\begin{figure*} 
  \epsscale{1.1}
  \plottwo{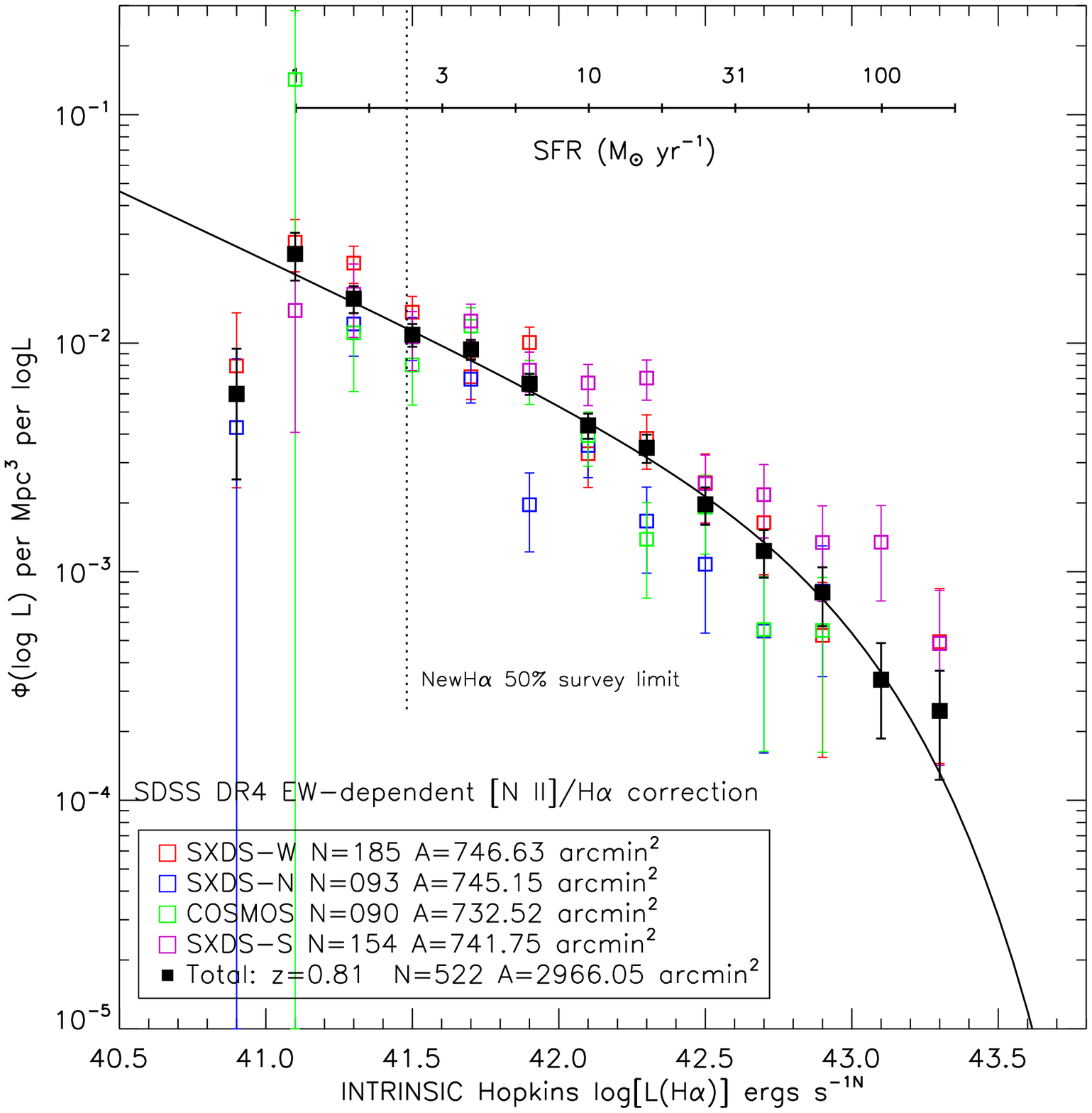}{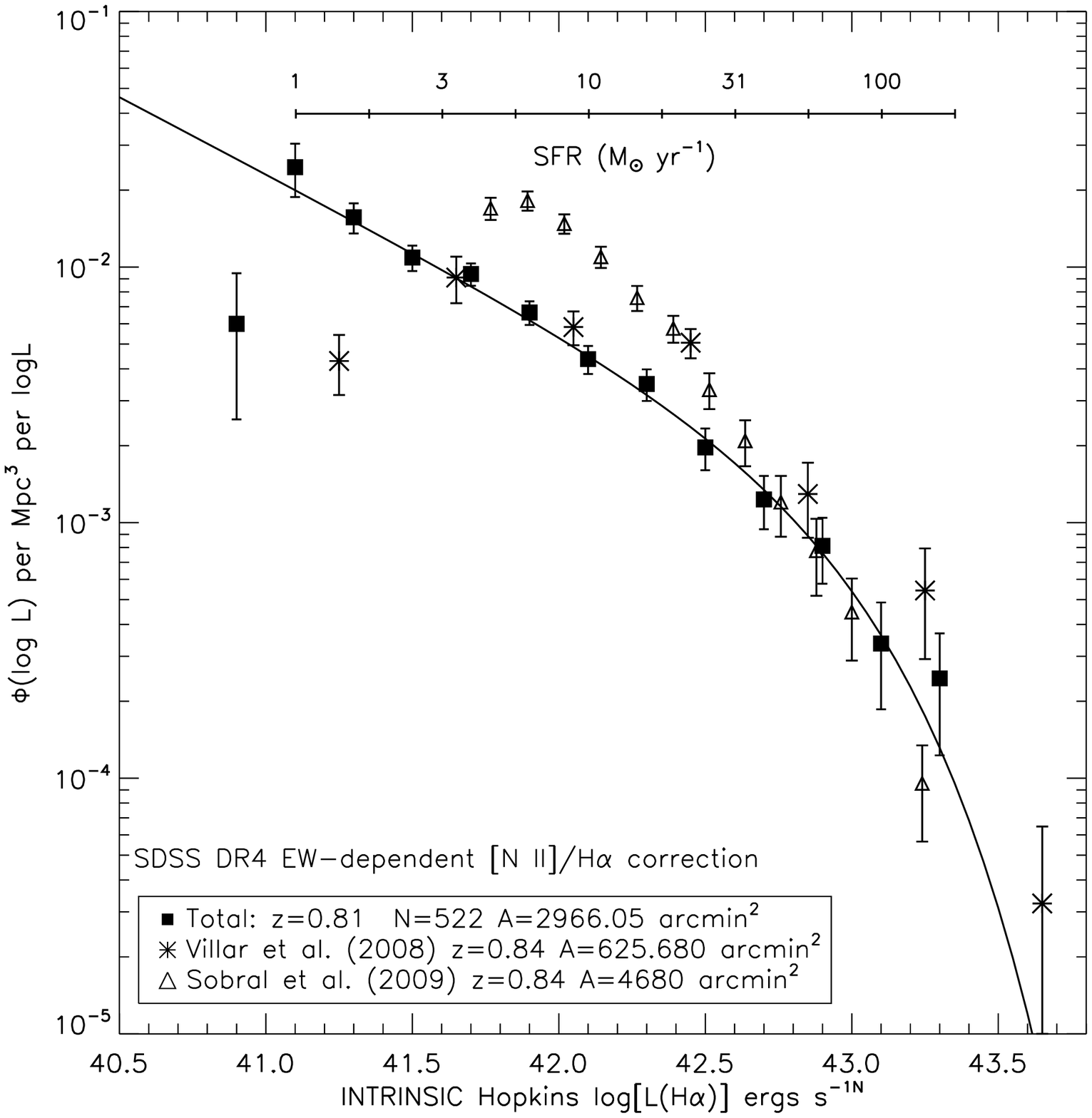}
  \caption{Extinction-corrected \Ha\ LF at $z\sim0.81$. Dust extinction corrections adopted the
    \cite{hopkins01} equation. The color and point-style conventions follow those in Figure~\ref{HaLF}. The left
    figure shows each of our NEWFIRM pointing and the average with the best-fitting Schechter function (\lstarp,
    \phistarp, and \alphap) as the solid line. On the right panel, we compare our average LF with the LFs of
    \citetalias{villar08} (shown as asterisks) and \citetalias{sobral09} (shown as triangles). \citetalias{sobral09}
    measurements were adjusted to adopt the \cite{hopkins01} dust extinction equation instead of A(\Ha) = 1.0 mag.
    Note that the ordinate axes have different ranges. Conversion between \Ha\ luminosities and SFRs is based
    on the \cite{kennicutt98} relation. (A color version of this figure is available in the online journal.)}
  \label{HaLFext}
\end{figure*}

Figure~\ref{HaLF} presents the (observed) \Ha\ LF for this survey with and without completeness
corrections. Applying the necessary completeness corrections and adopting the \cite{hopkins01} extinction correction, we
have the \Ha\ LF shown in Figure~\ref{HaLFext}. Both raw, and extinction- and completeness-corrected number densities
as a function of luminosity are listed in Table~\ref{table3}.
The binned LF can be summed together to obtain a model-free, lower-limit \Ha\ luminosity density of
\LDfbinb\ [\LDfbina].

\begin{figure} 
  \epsscale{1.1}
  \plotone{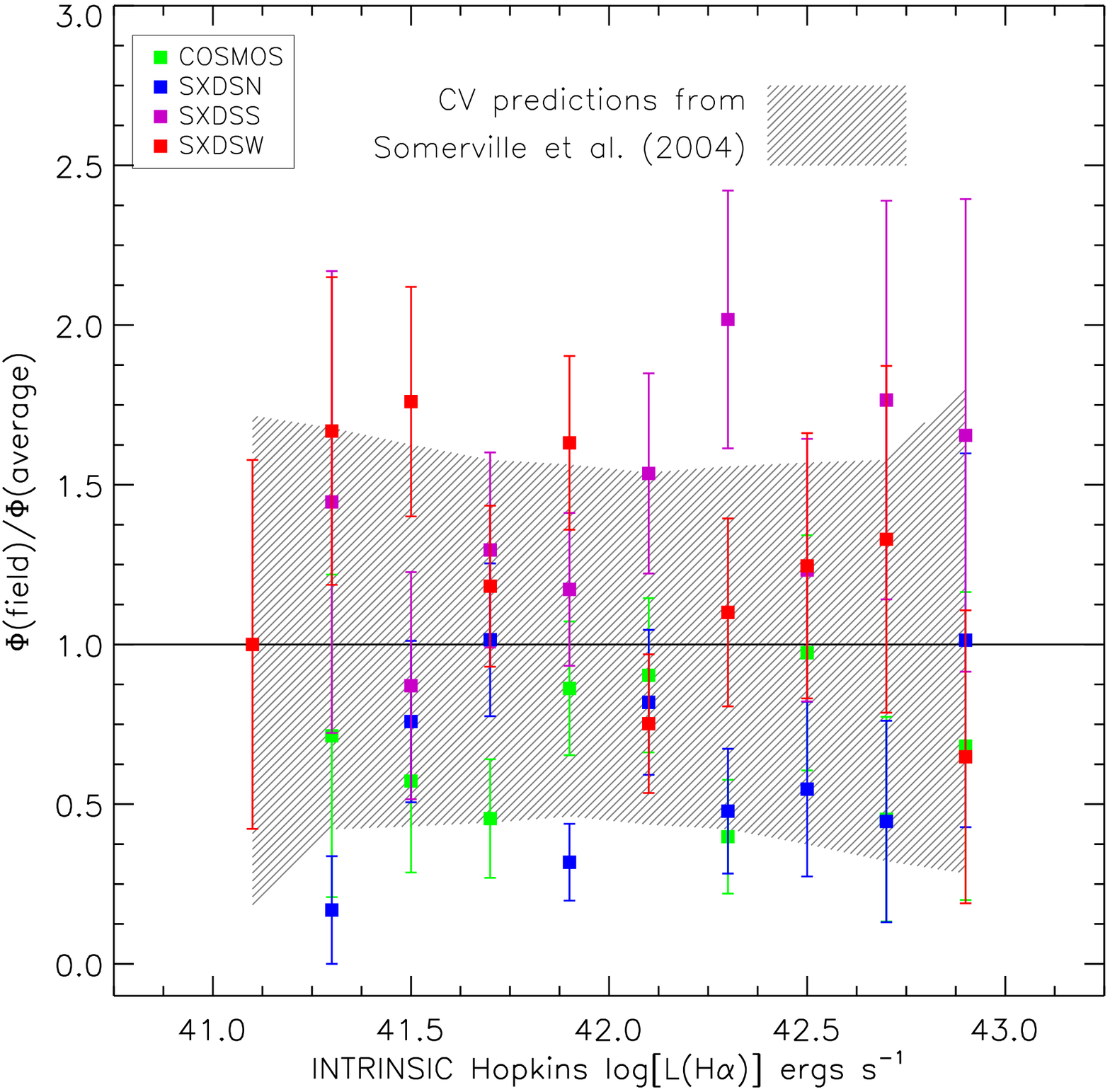}
  \caption{Field-to-field fluctuations in the \Ha\ LF for the NEWFIRM pointings. The $x$-axis shows the
    extinction-corrected \Ha\ luminosity while the $y$-axis shows the number density of sources normalized to the
    average. Color convention of points follows those used in Figure~\ref{HaLF}. The shaded regions represent the
    1$\sigma$ variation predicted from the $\Lambda$CDM model of \citetalias{somerville04}.
    (A color version of this figure is available in the online journal.)}
  \label{CV_LF}
\end{figure}

The median variation of the number density relative to the average of all four fields is $\sim$50\%, and we illustrate
in Figure~\ref{CV_LF} the fluctuation of the four different pointings relative to the average. Using predictions
from \citet[hereafter S04]{somerville04},\defcitealias{somerville04}{S04} we find that the expected fluctuation
per NEWFIRM pointing (shaded region in Figure~\ref{CV_LF}) is consistent with what is observed. The hourglass-like shape of
expected amount of field-to-field fluctuations is due to (1) the stronger clustering of luminous galaxies, (2) the
decrease of cosmic variance with number density, and (3) the small volume surveyed at low luminosities due to the
weakness of emission lines. Thus, the minimum of field-to-field variations is around an extinction-corrected \Ha\ 
luminosity of $1\times10^{42}$ erg s$^{-1}$.

\input tabLF_arxiv.tex

It is common to model the \Ha\ LF by fitting the LF with the \cite{schechter76} function:
\begin{equation}
  \Phi(L)dL =
  \phi_{\star}\left(\frac{L}{L_{\star}}\right)^{\alpha}\exp{\left(-\frac{L}{L_{\star}}\right)}\frac{dL}{L_{\star}}.
\end{equation}
This function was derived from the \cite{PS74} formalism, which describes the halo mass function,
and was adapted to explain the distribution of galaxy continuum luminosities.
An explanation for why the Schechter function can describe the continuum LF is that a connection exists between
the luminosities of galaxies and their stellar masses and/or halo masses. However, such a link may be much weaker
with the \Ha\ luminosity and SFR; thus, one might question whether the Schechter function is a good model for the
\Ha\ LF. For the current analysis we simply assume that the Schechter function can be used to adequately
model the distribution of \Ha\ luminosities and SFR since it appears to provide a good fit to our data. This also
facilitates comparisons with previous work. However, this assumption should be explored further when more
accurate LFs from future studies show evidence that a Schechter function is
not the best model to explain the \Ha\ LF.

\begin{figure} 
  \epsscale{1.1}
  \plotone{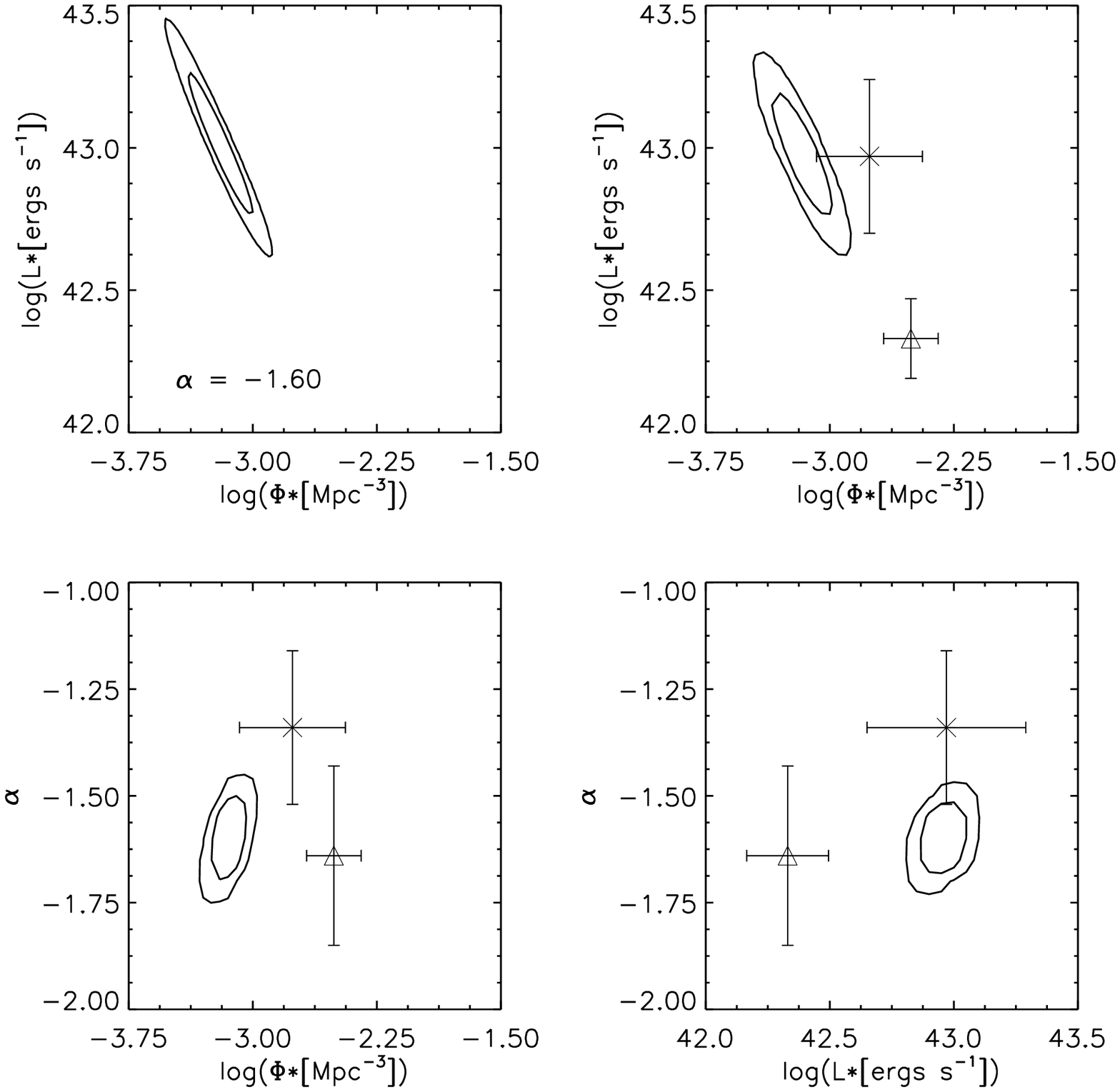}
  \caption{Confidence contours for the Schechter fit. 68\% and 95\% level for $\phistar$, $\Lstar$, and $\alpha$ are
    shown from a Monte Carlo simulation of the \Ha\ LF with extinction, completeness, and \NII\ contamination
    corrections. The faint-end slope is fixed to \alphaf\ for the upper left panel and is free for the other three
    panels. For these other three panels, we overlay the Schechter-fitting results of \citetalias{villar08} and
    \citetalias{sobral09} as asterisks and triangles, respectively.}
  \label{contours}
\end{figure}

In order to obtain the best-fitting Schechter parameters, a Monte Carlo simulation was performed to consider the
full range of scatter in the extinction- and completeness-corrected \Ha\ LF. We ignored luminosities below our
50\% completeness limit (\Llim\ observed; \Llimext\ extinction-corrected). Each datapoint was perturbed randomly
$1\times10^5$ times following a Gaussian distribution with $1\sigma$ in $\Phi(L)$ given by Poisson statistics. Each
iteration is then fitted to obtain the Schechter parameters. The best-fitting Schechter parameters are 
then determined from the averages of these fits. The confidence contours for the best fit are shown in
Figure~\ref{contours}.
A summary of the best fits and the corresponding integrated \Ha\ luminosity density ($\mathcal{L} = \int L\Phi(L) dL$)
and SFR density (see below) is provided in Table~\ref{LFtable}.
We find a relatively steep faint-end slope for the \Ha\ LF ($\alpha = \alphafa$) at $z\sim0.81$, indicating that
galaxies below $0.2$ and $1\Lstar$ contribute 61\% and 89\% to the total \Ha\ luminosity/SFR density, respectively.
Often, past studies have opted to fixed the faint-end slope since they were unable to reliably constrain it. Following
this methodology, we also report the results of our fits when $\alpha$ is set to $-$1.6 in Table~\ref{LFtable}.

The extinction-corrected \Ha\ luminosity density can be converted into a SFR density by using the recipe given in
\cite{kennicutt98}: SFR(\Ha) = $7.9\times10^{-42}L({\rm H}\alpha)$ where the SFR is given in \Msun\ \iyr\ and the
\Ha\ luminosity is given in erg s$^{-1}$. This conversion assumes a Salpeter IMF with minimum and maximum masses
of 0.1 \Msun\ and 100 \Msun\ and solar metallicity. We determined that the \Ha\ SFR density is \SFRp\ down to $L=0$,
where the second sets of errors account for cosmic variance estimated from \citetalias{somerville04}.
Compared to measurements at $z\lesssim0.1$ \citep{gallego95,perez03,brinchmann04,nakamura04,hanish06,ly07,westra10},
our \Ha\ SFR density at $z\sim0.8$ is higher by a factor of 3.8 to 16.6 with a median of 8.1.

It is thought that the \Ha\ luminosity density is dominated by emission from star formation and not active galactic nuclei (AGNs).
However, to accurately compute the SFR volume density, we statistically account for the fraction of the
\Ha\ luminosity volume density originating from AGNs.
While estimating the AGN fraction is an observational challenge because deep x-ray data 
and/or spectroscopic information are needed, previous studies have typically found that 10\%--15\% of
galaxies have AGNs. For example, \cite{brinchmann04} estimate that 11\% of the \Ha\ luminosity density
is due to AGNs at $z\lesssim0.1$. \cite{gallego95} found 15\% AGN contamination to the \Ha\ SFR density at $z\sim0$.
\citetalias{villar08} used X-ray data for a small ($\sim$50) sample of \Ha\ emitters at
$z\sim0.8$ and found 10\%$\pm$3\%. \citetalias{sobral09} looked at the \OIII/\Hb\ and \OII/\Hb\ flux ratios
\citep{rola97}, for a subset of 28 $z\sim0.8$ \Ha\ emitters with optical spectroscopy and reported 15\%$\pm$8\%. 
Our preliminary analysis of the rest-frame optical emission-line flux ratios from our IMACS spectroscopy (see
Section \ref{3.2.1}) finds similar results.  Based on the \OIII/\Hb\ and \OII/\Hb\ flux ratios, 5 (10) of 141 \Ha\
emitters are AGNs (LINERs). Thus, we correct the above \Ha\ SFR density by \AGNf\ to account for AGN and LINER
contamination. This reduces our total \Ha\ SFR density to \SFRpAGN.


\section{Comparisons with Other Near-Infrared \Ha\ Studies}\label{5}
Recently, two other independent groups have performed relatively wide-field near-infrared narrowband imaging on 3--4
m class telescopes to search for high-$z$ emission-line galaxies. \citetalias{villar08} first surveyed 626
arcmin$^2$ for \Ha\ emitting galaxies at $z\sim0.84$. They identified 165 galaxies and obtained a extinction-corrected
SFR density of 0.17 \Msun\ \iyr\ \vMpc. \citetalias{sobral09} surveyed a total of 1.3 deg$^2$, identified 743 \Ha\
emitting galaxies at $z\sim0.84$, and determined a SFR density of 0.1 \Msun\ \iyr\ \vMpc. New\Ha\ complements
these surveys through a combination of depth and volume surveyed: it covers almost 5 times more area than
\citetalias{villar08}, and while \citetalias{sobral09} covers about 50\% more area, our survey is 0.6 dex deeper.
These advantages simultaneously allow us to (1) obtain better constraints on the faint-end slope and
``knee'' of the LF (we acquired $\sim$10\% accuracy on the slope and 0.5 dex on $\Lstar$; see above),
and to (2) reduce field-to-field fluctuations.
The New\Ha\ dataset not only enables us to compute a more robust estimate of the LF, it also allows us to better
understand the properties of sub-$\Lstar$ galaxies and their role in the overall star formation history
of the universe.
Comparisons of the LFs between these three surveys are provided in
Figures~\ref{HaLF}, \ref{HaLFext}, and \ref{contours} and Table~\ref{LFtable}.
We begin by comparing how NB excess emitters are selected and how \Ha\ emitters are identified, and then discuss
the discrepancies between these surveys that are apparent.

{\it Redshift and sensitivity.}
The \citetalias{villar08} and \citetalias{sobral09} surveys probed volumes at a slightly higher redshift of $\sim0.84$.
Of course, significant evolution between $z\sim0.81$ and $z\sim0.84$ (corresponding to $\Delta t = 127$ Myr at $\sim6.5$
Gyr) is not expected. The main consequence of the different surveyed redshift arises from impact of sky background
level at the wavelength of redshifted \Ha.
New\Ha\ targets a cleaner window in the sky spectrum. This partly leads to the factor of two and four times deeper
depth in emission-line flux sensitivity compared to \citetalias{villar08} and \citetalias{sobral09}.

{\it Selection of excess emitters.}
All three surveys identify NB excess emitters above $2.5\sigma$ significance in the $J-$NB color
(though we also report results based on a more robust $3\sigma$-selected sample).
However, there are differences in the minimum $J-$NB color criterion and the aperture(s) used. For example, we
required at least a $\Delta$($J-$NB118)
color of 0.2 mag and used two aperture sizes (2\arcsec or 2.5\arcsec\ and 3 or 4\arcsec). \citetalias{sobral09} used a \aper{3}
selection and required a minimum $J-{\rm NB}$ color of 0.3 mag. \citetalias{villar08} indicated that they selected
sources in a total of 10 apertures out to 5 times the FWHM with a minimum $J-{\rm NB}$ color of 0.15 mag. We found
that the inclusion of larger apertures for New\Ha\ only provided 10\% more candidates that the smaller aperture failed
to catch. Larger apertures would only allow \citetalias{villar08} to identify bright and extended galaxies, which are
often at lower redshifts.

{\it Identification.}
In all three surveys, spectroscopic redshift is available for some NB excess emitters to distinguish other
emission lines from \Ha. The follow-up spectroscopy that we have classifies
\Pspecs\ [\Pspec] of our 3$\sigma$ [2.5$\sigma$] NB excess emitter candidates.
On the other hand, \citetalias{villar08} and \citetalias{sobral09} classified the majority of their NB
excess emitters with photometric redshift.
These surveys have follow-up spectroscopy for 9\% \citepalias[138/1527;][]{sobral09} and 48\% \citepalias[69/165;][]{villar08}.
It is difficult to compare directly their spectroscopic completeness against ours since (1) \citetalias{sobral09} did
not report the size of the UDS spectroscopic sample, thus the 9\% is a lower limit on their spectroscopic completeness,
and (2) \citetalias{villar08} only reported spectroscopy for their \Ha\ emitters rather than the full NB excess emitter
sample. For the latter, New\Ha\ has 52\% spectroscopic completeness of \Ha\ emitters. Nevertheless, the spectroscopic
completeness of New\Ha\ is higher than those of \citetalias{villar08} and \citetalias{sobral09}.

{\it Similarities in the LF.}
The observed number density of \Ha\ emitters is shown in Figure~\ref{HaLF} (left panel). It illustrates that all three
surveys agree to within $\sim$50\%, which indicates that the differences in the selection of \Ha\ emitters and NB excess
emitters do not significantly affect the observed number densities of galaxies. The \citetalias{villar08} observed LF
is generally higher, but it is consistent with our SXDS-S observations, which has a similar area coverage, so cosmic
variance is likely the cause (see below for further discussion).

In Figure~\ref{HaLFext} (right panel) we show the extinction- and completeness-corrected LF for all three \Ha\
surveys. It is apparent that the bright end of \citetalias{sobral09}'s LF is consistent with New\Ha's. This is to
be expected since both surveys cover large areas and incompleteness corrections are less of an issue. In addition,
\citetalias{villar08} is in agreement with New\Ha\ below a luminosity of $\sim$10$^{42}$ erg s$^{-1}$.
This is also expected since \citetalias{villar08} reaches comparable sensitivity to our survey. However, there are
two discrepancies worth addressing.

{\it Differences in the LF.}
First, the extinction- and completeness-corrected LF of \citetalias{villar08} is 0.2--0.3 dex higher at $\Lstar$.
However, the luminous end of the \citetalias{villar08} LF is not well determined since their survey consists of three
pointings (two in the Groth strip, one in GOODS-N) totaling less than 0.2 deg$^2$. Thus, the combination of
Poisson fluctuations and cosmic variance can explain the higher number density at the bright end.

The second discrepancy is that \citetalias{sobral09} find a higher number density of faint \Ha\ emitters, after  
survey completeness corrections are applied (see Figure~\ref{HaLF}, right panel). \citetalias{sobral09} claimed
that the completeness correction is a factor of 2--3 at their 2.5$\sigma$ flux limit, while our Monte Carlo simulation
indicates that 50\% completeness occurs at our $\sim6\sigma$ flux limit. Furthermore, \citetalias{sobral09} find
that their completeness gradually decline while our Monte Carlo simulation shows that above an observed \Ha\
emission-line luminosity of $\sim10^{41.5}$ erg s$^{-1}$, the completeness is generally $\gtrsim$80\%
and falls of rapidly with fainter luminosities. How rapidly the completeness declines will affect the shape of
the LF, and thus the completeness corrections can alter the determined Schechter parameters. 
An illustration of this is provided in Figure~\ref{contours} where the differences in the Schechter
parameters for these two studies are shown. Recall that the two LFs prior to any completeness corrections are
consistent.
  
The differences between New\Ha\ and \citetalias{sobral09} at the faint end imply that (1) deeper data must be
obtain such that incompleteness determinations are less of an issue or (2) a more standardized rigorous procedure
is needed for completeness estimates. For example, a method that considers a maximum likelihood approach that
simultaneously produces the observed emission-line EW and LF (such as those performed in Section \ref{3.1}).

\section{Discussion}\label{6}
\begin{figure}[htc] 
  \epsscale{1.1}
  \plotone{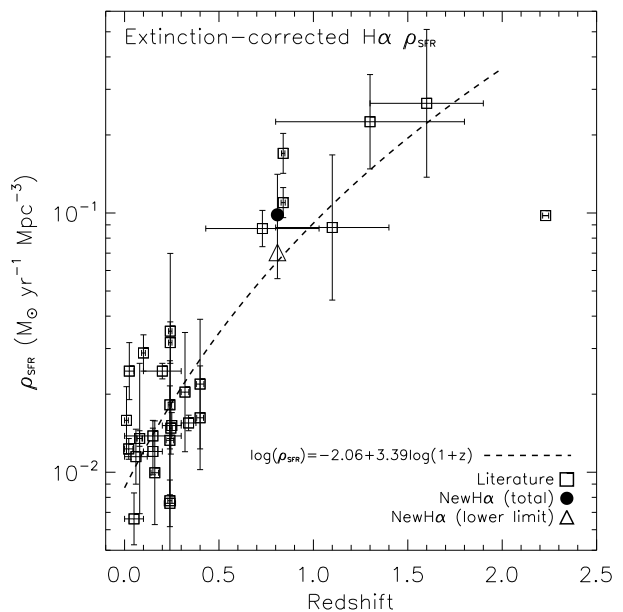}
  \caption{SFR density from \Ha\ surveys. Open squares are 31 measurements from the literature
    (see Section \ref{6} and Table~\ref{table4} for references), while New\Ha\ measurements are shown as the filled
    circle (LF integrated to $L=0$) and as a triangle (above the survey limit). Our \Ha\ measurements have been
    systematically reduced by 11\% to account for potential AGN contamination. The uncertainties in our ``total''
    SFR density include an estimate of the amount of cosmic variance expected for our survey (see text) and the
    uncertainties in fitting the LF with a Schechter profile. The dashed line is the fit adopted by \cite{dale10}
    for $z<2$ measurements (this fit excluded our measurement):
    $\log{\left(\SFRd[M_{\sun}\ {\rm yr}^{-1}]\right)} = -2.06 + 3.39\log{(1+z)}$.}
  \label{HaSFR}
\end{figure}
In this section, we compare our \Ha\ SFR density measurements with those published in the literature for a range of
redshifts. We limit the comparison to other \Ha-based measurements to avoid systematic issues with other SFR
indicators.

The latest compilation of \Ha\ measurements was made by \cite{dale10}. We add our measurements to this compilation
and plot them as a function of redshift in Figure~\ref{HaSFR}. All of the measurements plotted are summarized in
Table~\ref{table4}. We note that the measurements reported here correct for a few mistakes found in the
  original papers (see Table~\ref{table4} footnotes). The dashed line is a fit determined by \cite{dale10} where
the New\Ha\ SFR density was not included in the fitting process. It has the form of
$\log{\left(\frac{\SFRd}{M_{\sun}\ {\rm yr}^{-1}}\right)} = -2.06 + 3.39\log{(1+z)}$. Our SFR density measurement
with the removal of \AGNf\ for AGN contamination is above the \cite{dale10} fit, but consistent within
the uncertainties. This relation indicates that the \Ha\ SFR density increases by a factor of $\sim10$ per unit
redshift at $z<1.5$.
\input tab4_arxiv.tex

\begin{figure}[htc] 
  \epsscale{1.1}
  \plotone{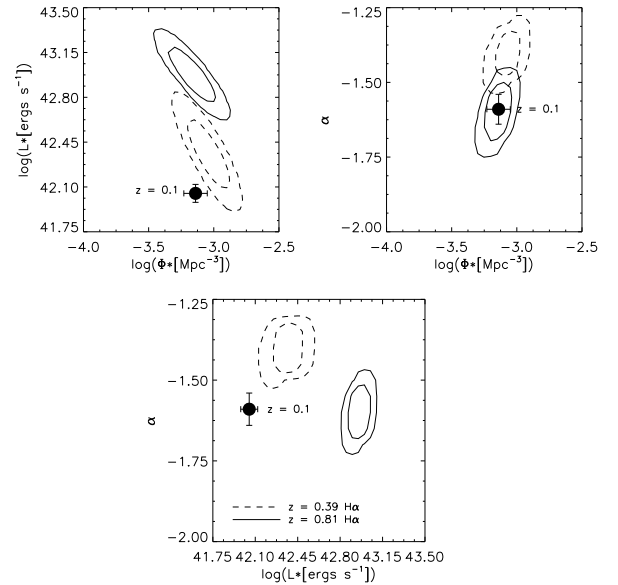}
  \caption{Redshift evolution in Schechter parameters. Confidence contours for $\Lstar$, $\phistar$, and $\alpha$
    for \Ha\ measurements at $z\sim0.1$ (black circles), $z=0.39$ (dashed line), and $z=0.81$ (solid line). These
    contours are derived from a Monte Carlo realizations of the \Ha\ LFs. The greatest difference is in $\Lstar$ with
    it increasing by $\sim1$ dex.}
  \label{schechter_evol}
\end{figure}

To understand this redshift evolution, we compare in Figure~\ref{schechter_evol} the confidence contours of the
Schechter parameters for measurements at $z=0.07$, 0.39, and 0.81.
The values reported for $z=0.39$ is based on a combination of data from two complementary surveys:
the Subaru Deep Field \citep[SDF;][extremely deep for 0.25 deg$^2$]{ly07} and
the Wyoming Survey for \Ha\ \citep[shallower sensitivity by 1.8 dex but covers $\sim$1 deg$^2$]{dale10}.
The binned LFs are combined together and fit with a Schechter function.
Likewise, $z<0.1$ measurements from the SDF are combined with \cite{gallego95}. We find that the characteristic luminosity
systematically increases by 0.95 dex (0.7 dex) from $z\sim0.1$ ($z\approx0.4$) to $z\approx0.8$, while the
normalization is similar at all three redshifts. This indicates that the increase in the SFR density is a result
of $z\sim0.8$ $\Lstar$ galaxies producing stars at a rate that is $\approx$10 times that of local $\Lstar$
($\sim10^{42.0}$ erg s$^{-1}$) galaxies.

Infrared surveys have provided complementary constraints on the evolution of the LF and SFR densities. These
measurements are sensitive to UV radiation of massive stars that are absorbed by dust and re-radiated at
rest wavelengths of 10--100 \mm. \cite{lefloch05} examined the infrared luminosity of 24\mm-selected galaxies and
found that for $z\lesssim1$, the LF evolves as $\log{\left(L_{\star,{\rm IR}}\right)} \propto 3.2^{+0.7}_{-0.2}(1+z)$ and
$\log{\left(\phi_{\star,{\rm IR}}\right)} \propto 0.7^{+0.2}_{-0.6}(1+z)$. These results of
significant $\Lstar$ evolution and weak $\phistar$ evolution are consistent with those of New\Ha,
granted this can be for a different population of galaxies.

In addition, we find that our \Ha\ SFR density measurement at $z\approx0.8$ is consistent with $z\sim1$ UV and
\OII\ SFR density measurements \citep[and references therein]{hopkins04}, though the scatter among the measurements
spans a factor of $\sim2$. The causes
for the large scatter include (1) cosmic variance, (2) systematic issues involving the SFR indicators, and (3)
the different selection biases that affect each of the surveys.
These issues also affect the LF and SFR density derived from infrared surveys.
All of these issues will be addressed over the next few years to improve the accuracy that the cosmic star formation
history can be determined. In general, cosmic variance will gradually become less of an issue with surveys covering at
least several deg$^2$.
In future work, data from the New\Ha\ survey, in combination with UV data from {\it GALEX}, mid-infrared fluxes
from {\it Spitzer}, and \OII\ fluxes from follow-up spectroscopy (see Section \ref{3.2.1}), will directly enable us to
compare the SFRs based on these indicators in hundreds of individual galaxies to address point (2).

And finally, upcoming studies, including ones based upon the New\Ha\ Survey, will investigate the differences in galaxy samples
that result from the use of different selection techniques.  For example, infrared surveys primarily probe the dustiest galaxies,
while UV surveys preferentially select the bluest galaxies, and emission-line selected surveys are biased against low-EW galaxies.
Direct comparisons of samples selected with different techniques in the same volumes will allow us to account for variations
in the SFR volume density, which are due to such selection biases. As mentioned in the Introduction, ultimately
it will also be possible to trace large fractions of the cosmic star formation history with a single indicator, and then
compare the consistently measured histories from multiple indicators.

\section{Conclusions}\label{8}
We have presented new measurements of the \Ha\ LF and SFR volume density
for galaxies at $z\sim0.8$, based on \NBlo\mm\ narrowband imaging from the New\Ha\ Survey.
With a $3\sigma$ \Ha\ emission-line flux depth of $\approx1.9\times 10^{-17}$ \fluxunit\ (a luminosity of
$\approx6\times10^{40}$ erg s$^{-1}$) and an area coverage of 0.82 deg$^2$, the New\Ha\ survey allows for a reduction
of field-to-field fluctuations to 10\%, and for robust estimates of the faint-end slope ($\sim$10\% accuracy) and the
location of the ``knee'' (0.5 dex accuracy) of the LF. We have identified  \NNBs\ NB excess emitters above 3$\sigma$,
and \NHas\ are classified as \Ha\ emission-line galaxies at $z\approx\zlo$. The classification utilized a large
spectroscopic sample providing unambiguous determination of redshifts for \Pspecs\ of the sample. These spectra were
also used to calibrate the multi-color selection of the remaining \Ha\ emitters without spectroscopic follow-up.

We constructed the extinction- and completeness-corrected \Ha\ LF. Corrections for \NII\ flux
contamination and the effective surveyed volume, as a function of line flux, were applied. The LF is well
described by a Schechter function with \lstarp, \phistarp, and \alphap.
When the LF is integrated to $L=0$, we determine a SFR density of \SFRpAGN. This SFR density is (on average) 8.1
times higher than $z\lesssim0.1$ measurements. We determined that the characteristic \Ha\ luminosity is systematically
higher at $z\sim0.81$ by 0.70 and 0.95 dex compared to $z=0.39$ and $z\sim0.1$ estimates, respectively.
The normalization of the LF at $z\sim0.8$ is similar to what is seen for $z\sim0$. This may imply that the cause of the
redshift evolution in the \Ha\ SFR density is a result of $z\sim0.8$ $\Lstar$ galaxies producing stars at a rate that
is $\approx$10 times that of typical galaxies seen locally.

The depth and completeness of current high-$z$ \Ha\ surveys significantly limit the accuracy to which 
the (1) \Ha\ SFR density, (2) the shape of the LF, and (3) its evolution can be measured. This is underscored by the
fact that three independent \Ha\ surveys (including New\Ha) show excellent agreement in the observed number densities,
but exhibit discrepancies at the factor of two level after completeness corrections are applied. These discrepancies
lead to different conclusions on the evolution of star-forming galaxies: while New\Ha\ shows redshift evolution
in $\Lstar$, \citetalias{villar08} report evolution in both $\Lstar$ and $\phistar$, and \citetalias{sobral09} find
evolution in $\phistar$.
The differences between the results presented here and that of \citetalias{villar08} can be attributed to cosmic
variance and small number statistics at the luminous end. Fully understanding the differences between our results
and that of \citetalias{sobral09} require a more in-depth comparison of the completeness corrections for both studies.
Future surveys probing fainter luminosities, which will circumvent uncertain completeness corrections
above emission-line fluxes of $\sim 2 \times 10^{-17}$ erg s$^{-1}$ cm$^{-2}$, in combination with rigorous simulations
for incompleteness that are consistently applied across studies, will allow for convergence on the true cosmic SFR
volume density, and better understand the factors that drive the evolution.

The OH background is the primary limitation for ground-based near-infrared surveys, which explains the historical dearth
of \Ha\ measurements at $z=0.5$--0.8 and in the $H$-band window ($z\sim1.5$). These epochs will be studied with
the new {\it Hubble}/WFC3 infrared grism. For example, \cite{atek10} has begun a survey, that will
yield \Ha\ SFRs at $z=0.25$--1.6 as well as other emission lines (e.g., \OII\ and \OIII) out to $z\sim4$ to probe a
significant fraction of the early universe. Both the NB and grism surveys complement one another through a combination of
surveyed area, depth, and redshift.

This study illustrates that \Ha\ can be extended to high redshift, and with more sensitive detectors and wide
field coverage in the near future, \Ha\ measurements for thousands of galaxies at $z\approx1$--3 will be
possible to trace cosmic star formation history with a consistent SFR indicator over the past 11
billion years.

\acknowledgements
We are grateful to Ron Probst, Buell Januzzi, and Ron George for their work to enable regular filter changes in NEWFIRM,
without which the New\Ha\ survey would not have been possible.
We thank Ivo Labbe for sharing his IDL processing pipeline
\citep[written for the reduction of data from the NEWFIRM Medium-Band Survey;][]{dokkum09},
which allowed for an initial reduction of our data in advance of the development of our dedicated
pipeline.
We also thank Hisanori Furusawa for providing his proprietary photometric redshift catalog for the SXDS field.
We thank Masami Ouchi and collaborators for including New\Ha\ narrowband excess objects
as mask fillers in their IMACS spectroscopic observations, and providing the data for these
$\sim$100 objects.
C.L. has been supported by NASA grant NNX08AW14H through their Graduate Student
Researcher Program.
The New\Ha\ Survey has been primarily funded by Hubble and Carnegie Fellowships to JCL.
This work has used zCOSMOS observations carried out using the Very Large Telescope at the ESO Paranal
Observatory under Programme ID: LP175.A-0839.
We thank the anonymous referee for their prompt response and helpful comments that improved the paper.
We also thank David Sobral, Jim Geach, and Philip Best for discussions about their \Ha\ LFs.

{\it Facilities:} \facility{Magellan:Baade (IMACS)}, \facility{Mayall (NEWFIRM)}, \facility{Subaru (Suprime-Cam)}, 
\facility{VLT:Melipal (VIMOS)}

\end{document}

%% file: tab1_arxiv.tex
\begin{deluxetable*}{lclccccccc}
  \tablewidth{0pt}
  \tabletypesize{\scriptsize}
  \tablewidth{0pc}
  \tablecaption{Summary of NEWFIRM Imaging}
  \tablehead{
    \colhead{Field} & \colhead{R.A., Dec} & \colhead{Observation Dates} & \colhead{Filter} & \colhead{Int. Time} &
    \colhead{FWHM} & \multicolumn{4}{c}{Limiting Magnitude (AB, 3$\sigma$)}\\
    \cline{7-10}
    & (J2000) & & &
    \colhead{(hr)}&
    \colhead{(arcsec)} &
    \colhead{SE} &
    \colhead{SW} &
    \colhead{NE} &
    \colhead{NW} 
  }
  \startdata
  COSMOS & 10:01, +02:01   & 2007 Dec 4,5                &    J &  2.30 & 1.20 & 23.63 & 23.71 & 23.74 & 23.74\\
         &                 & 2007 Dec 2-5                &NB118 &  8.16 & 1.50 & 23.48 & 23.54 & 23.55 & 23.59\\
  SXDS-N & 02:18, $-$04:38 & 2008 Sep 28,29, Oct 1,22,23 &    J &  3.52 & 1.10 & 23.91 & 24.03 & 24.03 & 24.05\\
         &                 & 2008 Sep 29, Oct 1,22,23    &NB118 &  8.47 & 1.20 & 23.77 & 23.94 & 23.90 & 23.98\\
  SXDS-S & 02:18, $-$05:15 & 2007 Dec 3-5                &    J &  2.40 & 1.25 & 23.67 & 23.75 & 23.81 & 23.78\\
         &                 & 2007 Dec 2-5                &NB118 & 10.28 & 1.60 & 23.40 & 23.51 & 23.58 & 23.61\\
  SXDS-W & 02:16, $-$04:57 & 2008 Sep 23-26,28, Oct 21,22&    J &  3.97 & 1.20 & 24.02 & 24.14 & 24.14 & 24.19\\
         &                 & 2008 Sep 23-26,28, Oct 21,22&NB118 & 12.67 & 1.15 & 23.88 & 23.96 & 24.06 & 24.03\\[-3mm]
\enddata
\label{table1}
\tablecomments{Limiting magnitudes are reported for apertures sizes of 2\farcs0 diameter for the SXDS-N and SXDS-W,
and 2\farcs5 for the SXDS-S and COSMOS.  These apertures are chosen to contain a minimum of 80\% of the flux 
of a point source for each pair of $J$ and NB118 imaging, given the seeing conditions during the observations.}
\end{deluxetable*}

%% file: tab2.v1.arxiv.tex
\begin{deluxetable*}{lccrrr}
\tablewidth{0pt}
\tabletypesize{\scriptsize}
\tablecaption{Summary of NB118 Selected Samples}
\tablehead{
  \colhead{Field} &
  \colhead{Area} &
  \colhead{$N_{\rm NB1187}$} &
  \colhead{$f_{\rm spec}$} &
  \colhead{$N_{H\alpha}$}&
  \colhead{$f_{\rm spec}$(\Ha)} 
}
\startdata

COSMOS &  732.5 & 157 [253]    & 65\% [48\%] &  59 [ 90]  & 46\% [32\%] \\
SXDS-N &  745.2 & 157 [255]    & 59\% [42\%] &  63 [ 93]  & 63\% [49\%] \\
SXDS-S &  741.7 & 201 [265]    & 75\% [64\%] & 130 [154]  & 76\% [70\%] \\
SXDS-W &  746.6 & 303 [445]    & 52\% [38\%] & 142 [185]  & 61\% [48\%] \\
Totals & 2966.0 & \NNBs [\NNB] & \Pspecs\ [\Pspec] & \NHas\ [\NHa] & 64\% [52\%] \\[-3mm]
\enddata
\label{table2}
\tablecomments{Surveyed area listed in units of arcmin$^2$. $N_{\rm NB1187}$ gives the number of
  sources selected as emission-line galaxy candidates which have 3$\sigma$ [2.5$\sigma$]
  $\Delta(J-{\rm NB118})$ excess and are above $\Delta(J-{\rm NB118})=0.2$ mag.  $f_{\rm spec}$ indicates the percentage of N$_{\rm NB1187}$
  with follow-up optical spectroscopy.  $N_{H\alpha}$ gives the number of emission-line galaxy
  candidates identified as \Ha\ emitters by empirical color selection or spectroscopy, as
  discussed in Section \ref{2.1}.  $f_{\rm spec}$(\Ha) indicates the percentage of
  $N_{H\alpha}$ that has been spectroscopically confirmed.}
\end{deluxetable*}

%% file: tab3_arxiv.tex
\begin{deluxetable*}{rrrlccrrlc}
\tablewidth{0pt}
\tabletypesize{\scriptsize}
\tablecaption{\Ha\ Luminosity Function at $z\sim0.81$}
\tablehead{
  & 
  \multicolumn{4}{c}{2.5$\sigma$} & &
  \multicolumn{4}{c}{3$\sigma$}\\
  \cline{2-5} \cline{7-10}
  \colhead{$\log{L}$} & 
  \colhead{$N$} &
  \colhead{$N_{\rm spec}$} &
  \colhead{$\Phi(L)$} &
  \colhead{\C} & 
  &
  \colhead{$N$} &
  \colhead{$N_{\rm spec}$} &
  \colhead{$\Phi(L)$} &
  \colhead{\C}
}
\startdata
\multicolumn{10}{c}{Raw number densities}\\
40.70  &   11  &    0 &    11.17$\pm$3.4  & 0.08 & &\ldots&\ldots&    \ldots        &\ldots\\
40.90  &   60  &    8 &    49.34$\pm$6.4  & 0.22 & &   20 &    3 &    12.74$\pm$2.8 &  0.22\\
41.10  &  116  &   27 &    83.76$\pm$7.8  & 0.47 & &   65 &   19 &    40.36$\pm$5.0 &  0.49\\
41.30  &  102  &   50 &    63.02$\pm$6.2  & 0.72 & &   78 &   45 &    45.96$\pm$5.2 &  0.73\\
41.50  &   98  &   74 &    55.39$\pm$5.6  & 0.79 & &   96 &   73 &    54.01$\pm$5.5 &  0.79\\
41.70  &   66  &   54 &    36.21$\pm$4.5  & 0.79 & &   66 &   54 &    36.21$\pm$4.5 &  0.79\\
41.90  &   34  &   26 &    18.65$\pm$3.2  & 0.81 & &   34 &   26 &    18.65$\pm$3.2 &  0.81\\
42.10  &   19  &   18 &    10.42$\pm$2.4  & 0.80 & &   19 &   18 &    10.42$\pm$2.4 &  0.80\\
42.30  &    8  &    7 &     4.39$\pm$1.6  & 0.81 & &    8 &    7 &     4.39$\pm$1.6 &  0.81\\
42.50  &    8  &    8 &     4.39$\pm$1.6  & 0.86 & &    8 &    8 &     4.39$\pm$1.6 &  0.86\\\hline
\multicolumn{10}{c}{Extinction and completeness-corrected number densities}\\
40.90  &    3  &    0 &    59.79$\pm$34.5 & 0.07 & &\ldots&\ldots&    \ldots         &\ldots\\
41.10  &   18  &    1 &   245.40$\pm$57.8 & 0.14 & &    3 &   0  &    43.48$\pm$25.1 & 0.18\\
41.30  &   55  &    8 &   156.58$\pm$21.1 & 0.28 & &   19 &   4  &    46.46$\pm$10.7 & 0.26\\
41.50  &   77  &   15 &   108.82$\pm$12.4 & 0.52 & &   43 &  10  &    50.12$\pm$7.6  & 0.53\\
41.70  &   99  &   44 &    93.94$\pm$9.4  & 0.69 & &   64 &  37  &    54.58$\pm$6.8  & 0.71\\
41.90  &   89  &   55 &    66.39$\pm$7.0  & 0.78 & &   84 &  53  &    61.75$\pm$6.7  & 0.78\\
42.10  &   63  &   48 &    43.66$\pm$5.5  & 0.80 & &   63 &  48  &    43.66$\pm$5.5  & 0.80\\
42.30  &   50  &   43 &    34.85$\pm$4.9  & 0.79 & &   50 &  43  &    34.85$\pm$4.9  & 0.79\\
42.50  &   29  &   24 &    19.70$\pm$3.7  & 0.81 & &   29 &  24  &    19.70$\pm$3.7  & 0.81\\
42.70  &   18  &   14 &    12.32$\pm$2.9  & 0.80 & &   18 &  14  &    12.32$\pm$2.9  & 0.80\\
42.90  &   12  &   12 &     8.12$\pm$2.3  & 0.81 & &   12 &  12  &     8.12$\pm$2.3  & 0.81\\
43.10  &    5  &    4 &     3.37$\pm$1.5  & 0.81 & &    5 &   4  &     3.37$\pm$1.5  & 0.81\\
43.30  &    4  &    4 &     2.46$\pm$1.2  & 0.89 & &    4 &   4  &     2.46$\pm$1.2  & 0.89\\[-3mm]
\enddata
\label{table3}
\tablecomments{$\Phi(L)$ is normalized to $1\times10^{-4}$ \vMpc\ dex$^{-1}$ and luminosities ($L$) are
  given in erg s$^{-1}$. \C\ is the survey completeness fraction defined in Section \ref{3.1}.
  Numbers reported in the top half are prior to any completeness corrections
  while completeness is included for the bottom half. The spectroscopic completeness
  ($N_{\rm spec}$) as a function of luminosity is shown.}
\end{deluxetable*}

%% file: tabLF_arxiv.tex
\begin{deluxetable*}{ccccccccc}
  \tablewidth{0pt}
  \tablecaption{Schechter Fits, \Ha\ Luminosity Densities, and SFR Densities}
  \tablehead{
    \colhead{Survey} & 
    \colhead{$\log{\Lstar}$} &
    \colhead{$\log{\phistar}$} &
    \colhead{$\alpha$} &
    \colhead{$\log{\mathcal{L}} (0)$} &
    \colhead{$\log{\mathcal{L}} (L_{\rm lim})$} &
    \colhead{$\log{\SFRd} (0)$} &
    \colhead{$\log{\SFRd} (0)$\tablenotemark{a}} & 
    \colhead{$\log{\SFRd} (L_{\rm lim})$\tablenotemark{a}}
  }
  \startdata
  New\Ha\               & $\lstarpa$ & $\phistarpa$ & $\alphapa$ & $\LDpa$ & $\LDplima$ & $\SFRpa$ & $\SFRpAGNa$ & $\SFRplima$\\
  New\Ha\               & $\lstarfa$ & $\phistarfa$ & $\alphafa$ & $\LDfa$ & $\LDflima$ & $\SFRfa$ & $\SFRfAGNa$ & $\SFRflima$\\
  \citetalias{villar08} & $\VLs$     & $\VPs$       & $\Va$      & 40.35   & \ldots     & $-$0.76  & $-$0.80     & \ldots\\
  \citetalias{sobral09} & $\SLs$     & $\SPs$       & $\Sa$      & 40.21   & \ldots     & $-$0.89  & $-$0.96     & \ldots\\[-3mm]
  \enddata
  \tablenotetext{1}{Corrections for AGN contamination applied.}
  \tablecomments{$\Lstar$ and $\phistar$ in units of erg s$^{-1}$ and Mpc$^{-3}$, respectively. Luminosity ($\mathcal{L}$)
    and SFR densities ($\rho_{\rm SFR}$) are provided for $L\geq0$ and $L\geq L_{\rm lim}.$
    All LFs adopt \cite{hopkins01} dust attenuation.
    Note that \citetalias{sobral09} incorrectly reported the normalization ($\phistar$) for their \Ha\ LF and
    was off by a factor of $\sim$0.43. D. Sobral provided a new preliminary fit, which adopts the \cite{hopkins01}
    extinction and the proper LF normalization.}
  \label{LFtable}
\end{deluxetable*}

%% file: tab4_arxiv.tex
\newcommand\zerr[2]{#1$\pm$#2}%
\begin{deluxetable*}{lcrrc}
  \tablewidth{0pt}
  \tablecaption{Compilation of \Ha\ SFR Densities}
  \tablehead{
    \colhead{References} &
    \colhead{$z$} &
    \colhead{Area\tablenotemark{a}} &
    \colhead{$N$} &
    \colhead{$\log{(\SFRd)}$\tablenotemark{b}}
  }
  \startdata
  \cite{gallego95}                   & \zerr{0.022}{0.022} & 471.4 deg$^2$      &  176 & $-1.91\pm0.04$\\
  \cite{TM98}                        & \zerr{0.20}{0.10}   & 500                &  138 & $-1.61\pm0.03$\\ 
  \cite{yan99}                       & \zerr{1.3}{0.5}     &$\sim85$            &   33 & $-0.574\pm0.182$\\ 
  \cite{sullivan00}                  & \zerr{0.15}{0.15}   & \ldots             &  216 & $-1.86\pm0.06$\\
  \cite{tresse02}                    & \zerr{0.73}{0.30}   & \ldots             &   30 & $-1.06^{+0.07}_{-0.08}$\\
  \cite{fujita03}                    & \zerr{0.242}{0.009} & 706                &  348 & $-1.50^{+0.08}_{-0.17}$\tablenotemark{c}\\
  \cite{hippelein03}                 & \zerr{0.245}{0.007} & 407                &   92 & $-1.83^{+0.10}_{-0.13}$\\
  \cite{perez03}                     & \zerr{0.025}{0.025} & \ldots             &   79 & $-1.61^{+0.11}_{-0.08}$\\
  \cite{brinchmann04}                & \zerr{0.10}{0.01}   & SDSS               &\ldots& $-1.54\pm0.07$\\
  \cite{nakamura04}                  & \zerr{0.06}{0.06}   & SDSS               & 1482 & $-1.94^{+0.106}_{-0.082}$\\
                                     & \zerr{0.079}{0.013} & \ldots             &\ldots& $-1.87\pm0.03$\\
  \cite{hanish06}                    & \zerr{0.06}{0.06}   & SINGG              &  110 & $-1.80^{+0.13}_{-0.07}$\\
  \cite{ly07}                        & \zerr{0.08}{0.015}  & 868                &  318 & $-1.87\pm0.29$\tablenotemark{d}\\
                                     & \zerr{0.24}{0.011}  & 868                &  259 & $-2.11\pm0.24$\tablenotemark{d}\\
                                     & \zerr{0.40}{0.018}  & 868                &  391 & $-1.79\pm0.20$\tablenotemark{d}\\
  \cite{geach08}                     & \zerr{2.23}{0.016}  & 0.60 deg$^2$       &   55 & $-1.00$\tablenotemark{e}\\ 
  \cite{morioka08}                   & \zerr{0.242}{0.009} & 875+SDSS           & 575 & $-1.456^{+0.30}_{-0.174}$\\
  \cite{shioya08}                    & \zerr{0.24}{0.009}  & 5540               &  980 & $-1.74^{+0.17}_{-0.097}$\\
  {\bf \citetalias{villar08}}        & \zerr{0.84}{0.009}  & 625                &  165 & $-0.77\pm0.077$\\
  \cite{westra08}                    & \zerr{0.24}{0.03}   & 1771               &  707 & $-2.12^{+0.09}_{-0.12}$\\ 
  \cite{shim09}                      & \zerr{1.1}{0.3}     &$\sim104$           &   35 & $-1.056\pm0.28$\\
                                     & \zerr{1.6}{0.3}     &$\sim104$           &   45 & $-0.577\pm0.285$\\
  {\bf \citetalias{sobral09}}        & \zerr{0.84}{0.011}  &         1.3 deg$^2$&  743 & $-0.960$\tablenotemark{e}\\
  \cite{dale10}                      & \zerr{0.16}{0.02}   &        4.19 deg$^2$&  214 & $-2.002\pm0.20$\\ 
                                     & \zerr{0.24}{0.02}   &        4.03 deg$^2$&  424 & $-1.877\pm0.21$\\
                                     & \zerr{0.32}{0.02}   &        4.13 deg$^2$&  438 & $-1.691\pm0.23$\\
                                     & \zerr{0.40}{0.02}   &        1.11 deg$^2$&   91 & $-1.660\pm0.25$\\
  \cite{hayes10}                     & \zerr{2.19}{0.014}  &       56           &   55 & \ldots\tablenotemark{f}\\
  \cite{westra10}                    & \zerr{0.05}{0.05}   &          4 deg$^2$ &  322 & $-2.18\pm0.10$\\
                                     & \zerr{0.15}{0.05}   &          4 deg$^2$ & 1127 & $-1.92\pm0.09$\\
                                     & \zerr{0.25}{0.05}   &          4 deg$^2$ & 1268 & $-1.82\pm0.05$\\
                                     & \zerr{0.34}{0.04}   &          4 deg$^2$ &  848 & $-1.81\pm0.03$\\
  New\Ha\ (total)                    & \zerr{0.809}{0.008} &       0.82 deg$^2$ & \NHa & $-1.00\pm0.18$\tablenotemark{d}\\
  New\Ha\ ($L\geq L_{\rm lim})$      & \zerr{0.809}{0.008} &       0.82 deg$^2$ &  414 & $-1.10\pm0.09$\tablenotemark{d}\\[-3mm]
  \enddata
  \label{table4}
  \tablenotetext{1}{Unless otherwise indicated, areas are in arcmin$^2$.}
  \tablenotetext{2}{$\SFRd$ in units of \Msun\ \iyr\ \vMpc. Corrections for dust extinction have been included.
    These values integrated the LF to $L=0$ except for the last line in this table.}
  \tablenotetext{3}{See \cite{ly07} for discussion of potentially 50\% contamination, which was not accounted.}
  \tablenotetext{4}{Estimates for cosmic variance are included within the uncertainties.}
  \tablenotetext{5}{We determined that the normalization of the LF, $\phistar$, was reported incorrectly
      in these papers by a factor of $\sim0.43$. Here, we report the correct SFR densities.}
  \tablenotetext{6}{Since \cite{hayes10} used the LF of \cite{geach08} for constraints on the
    bright end, and the normalization of the \cite{geach08} LF was reported incorrectly, the reported
    SFR density of \cite{hayes10} is thus affected and not reported here.}
\end{deluxetable*}